\documentclass[a4paper,11pt]{article}
\pdfoutput=1 

\usepackage{jcappub} 

\usepackage[T1]{fontenc} 
\usepackage{xspace}
\usepackage{aas_macros}

\newcommand{\nn}{\nonumber}
\newcommand{\pow}[1]{\times 10^{#1}}
\newcommand{\uni}[2]{~{\rm #1}^{#2}} 
\newcommand{\eqn}[1]{eq.~(\ref{#1})}
\newcommand{\eqns}[2]{eqs.~(\ref{#1}) and (\ref{#2})}

\newcommand{\secn}[1]{section~\ref{#1}}
\newcommand{\appndx}[1]{appendix~\ref{#1}}
\newcommand{\fig}[1]{figure~\ref{#1}}

\newcommand{\tab}[1]{table~\ref{#1}}

\newcommand{\Mfid}{\textsf{Fiducial}\xspace}
\newcommand{\Mstat}{\textsf{Static}\xspace}
\newcommand{\Mtuned}{\textsf{Static-Tuned}\xspace}

\newcommand{\Msink}{\textsf{Sink Method}\xspace}
\newcommand{\Msrc}{\textsf{Source Method}\xspace}

\title{\textsf{SCRIPT} in the Cosmic Dawn: Distinguishing Redshift-Evolving Galaxy Populations with 21-cm Fluctuations}

\author[a, 1]{Janakee Raste,\note{Corresponding author.}}
\author[a]{Tirthankar Roy Choudhury}
 
\affiliation[a]{National Centre for Radio Astrophysics, TIFR,\\Pune University Campus, Ganeshkhind, Pune 411007, India}
 
\emailAdd{janakee.raste@gmail.com}
\emailAdd{tirth@ncra.tifr.res.in}

\abstract{Recent James Webb Space Telescope (JWST) observations suggest an unexpectedly high abundance of luminous galaxies at $z \gtrsim 10$, challenging traditional models of star formation. The 21-cm signal of neutral hydrogen offers a complementary, volume-averaged probe of these early epochs. In this work, we present a major extension to the explicitly photon-conserving semi-numerical framework \textsf{SCRIPT}, enabling the self-consistent calculation of spin temperature fluctuations driven by inhomogeneous Lyman-$\alpha$ coupling and X-ray heating during the Cosmic Dawn. Using this framework, we compare a JWST-informed fiducial galaxy model, featuring a redshift-evolving star-formation efficiency, against a static baseline model. We demonstrate that the global 21-cm signal is highly degenerate: a static galaxy population can replicate the absorption trough of an evolving population if a steep, mass-dependent X-ray efficiency is invoked to shift intergalactic medium heating toward highly abundant, low-mass halos. However, this degeneracy is definitively broken by the 21-cm power spectrum. Because spatial fluctuations retain a structural memory of the clustering bias of the source halos, the tuned static model predicts a substantially lower power spectrum amplitude across all characteristic peaks. We conclude that combining global and fluctuation measurements is essential to robustly constrain the redshift evolution of high-redshift galaxies.}

\begin{document}

\date{}

\maketitle
\flushbottom

\section{Introduction}\label{sec:intro}

Recent observations with the James Webb Space Telescope (JWST) have transformed our understanding of the high-redshift Universe ($z \gtrsim 10$), uncovering a population of galaxies that appears significantly more abundant and luminous than predicted by standard galaxy formation models \citep{Naidu2022, Castellano2022, Finkelstein2022, Labbe2023, Atek2023, Adams2023, Bradley2023, Whitler2023, Robertson2024, Castellano2023_UVLF, Finkelstein2023_UVLF, Gonzalez2023, Donnan2023, Harikane2023, Bouwens2023, McLeod2024, Adams2024, Finkelstein2024_UVLF, Whitler2025, Gonzalez2025, Weibel2025}. While traditional models successfully reproduce ultraviolet luminosity functions (UVLFs) at $z \sim 6$, various astrophysical mechanisms must be invoked to explain this emerging high-redshift excess. These include an enhanced efficiency or stochasticity of star formation \citep{Dekel2023, Li2024, Qin2023, Chakraborty2024, Chakraborty2026, Mirocha2023, Mason2023, Shen2023, Sun2023, Pallottini2023, Gelli2024, Kravtsov2024}, a top-heavy stellar initial mass function \citep{Inayoshi2022, Trinca2023, Harikane2023, Ventura2024, Yung2024, Hutter2025, Lu2025}, minimal dust attenuation at early times \citep{Ferrara2023, Ziparo2023, Ferrara2024, Ferrara2025}, or a significant contribution from accreting black holes \citep{Inayoshi2022, Pacucci2022, Hegde2024, Fujimoto2024}.

While JWST directly samples the bright end of the galaxy population at these redshifts, the neutral hydrogen ($\mathrm{HI}$) 21-cm signal from the Cosmic Dawn (CD) provides a complementary, volume-averaged view of the intergalactic medium (IGM). Several experiments, such as EDGES \cite{Bowman2018}, SARAS \cite{Singh2022}, REACH \cite{deLeraAcedo2022}, MIST \cite{Monsalve2024}, RHINO \cite{2025RASTI...4af046B}, and PRIZM \cite{Philip2019}, are targeting the sky-averaged global signal. Simultaneously, interferometers like GMRT \cite{Paciga2013}, MWA \cite{Tingay2013}, LOFAR \cite{vanHaarlem2013}, and HERA \cite{DeBoer2017} are probing spatial fluctuations via the power spectrum, paving the way for future tomographic mapping by the SKA \cite{2026arXiv260626435C}. Because direct UVLF measurements at the highest redshifts remain restricted to luminous systems and may be subject to sample variance or dust uncertainties, the 21-cm signal serves as a crucial, independent probe of the faint, abundant galaxy population that dominates the total cosmic photon budget. This signal directly encodes the history of Lyman-$\alpha$ coupling, X-ray heating, and partial ionization driven by sources below the detection thresholds of direct imaging (for reviews, see \citep{2006PhR...433..181F,Pritchard2012}). Consequently, both the trajectory of the global 21-cm signal and the scale dependence of its spatial fluctuations carry clear signatures of the integrated star formation history and the instantaneous galaxy population across cosmic time. Crucially, testing whether pre-JWST galaxy evolution models (which lack redshift evolution in star-formation efficiency) can be independently ruled out by 21-cm observables allows us to evaluate if the high-redshift galaxy abundance inferred by JWST extends down to the faintest star-forming halos \citep{Park++2019,Qin2020,Munoz2022,Nikolic2024,Bevins2024,Pochinda2024,Dhandha2025,Dhandha2025a,Davies2025}.  

To interpret 21-cm observations during the Cosmic Dawn, theoretical models must explicitly account for spatial fluctuations in the spin temperature $T_S$, which are governed by inhomogeneous Lyman-$\alpha$ coupling and X-ray heating. Implementing such effects in full radiative-transfer hydrodynamical simulations is extremely challenging due to their computational expense \cite{Semelin2007,Baek2009,Semelin2017}, driving a reliance on fast semi-numerical algorithms. These approaches employ distinct methodologies to balance physical accuracy with computational speed. For instance, frameworks based on the excursion set formalism, such as \textsf{21CMFAST} \citep{21CMFAST} and \textsf{SIMFAST21} \citep{Santos2010}, typically compute radiation backgrounds by filtering the density and source fields utilizing Fourier-space techniques. Similarly, \textsf{21CMSPACE} \citep{Fialkov2012,Visbal2012} evaluates IGM ionization via the excursion set formalism but handles radiative transfer by propagating pixel-level analytic emissivities through numerical window functions. An alternative strategy relies on one-dimensional radiative transfer methods, where codes like \textsf{BEARS} \citep{Thomas2009} and \textsf{GRIZZLY} \citep{Ghara2018} model IGM heating and coupling by painting spherically symmetric radiation profiles around individual dark matter halos. This methodology has recently been advanced by the \textsf{BEORN} framework \citep{BEORN}, which generates three-dimensional maps while self-consistently accounting for the time-evolution of individual sources—specifically, the cosmological redshifting of photons and the growth of source luminosity during emission. In addition to these simulation-based frameworks, analytical models provide complementary insight into the global 21-cm signal \cite{Mirocha14,ECHO21} and spatial fluctuations generated during the partially Lyman-$\alpha$-coupled and X-ray-heated phases \citep{RS18,RS19,Zeus21}. However, despite the computational efficiency of these semi-numerical and analytical methods, maintaining strict photon conservation, ensuring convergence across varying spatial and redshift resolutions, and avoiding grid-discretization artifacts remain critical challenges for fast parameter-space exploration.

In this work, we present a major extension of \textsf{SCRIPT} (Semi-numerical Code for Reionization with PhoTon-conservation), an explicitly photon-conserving framework previously developed for reionization studies \citep{SCRIPT,Choudhury:2025}. This update incorporates a full, self-consistent treatment of $T_S$-fluctuations, enabling \textsf{SCRIPT} to demonstrate how different JWST-consistent galaxy populations leave distinct, non-degenerate signatures in the 21-cm global signal and power spectrum. To achieve this, \textsf{SCRIPT} implements the radiative transfer of Lyman-$\alpha$ and X-ray photons through two independent numerical schemes: a source-centric and a sink-centric algorithm. Both methods are engineered to remain robust across a broad range of spatial and redshift resolutions, and their detailed numerical agreement provides an independent internal validation of our algorithms. This dual approach ensures the accurate modeling of the fluctuation epoch, even within the large-volume, coarse-resolution grids required for rapid parameter space exploration.

Using this updated framework, we investigate the 21-cm signatures predicted by JWST-informed galaxy formation models. To highlight the diagnostic power of Cosmic Dawn observables, we compare a \Mfid model, in which star-formation efficiency evolves with redshift to match JWST UVLFs across $6 \lesssim z \lesssim 13.2$, against a \Mstat (redshift-independent) model representative of traditional pre-JWST frameworks. While the \Mstat model is disfavored by bright-end JWST UVLFs at $z > 9$, adopting it as a reference baseline allows us to address a critical question: Can flexible Lyman-$\alpha$ and X-ray astrophysical parameters mimic the 21-cm signal of a redshift-dependent galaxy population? By attempting to force the \Mstat model to reproduce the fiducial global signal through highly tuned IGM heating and coupling parameters, we demonstrate that spatial fluctuations retain a structural memory of the underlying halo population. This effectively breaks astrophysical degeneracies that cannot be resolved by global-signal measurements alone.

This paper is organized as follows. In \secn{sec:method}, we describe the numerical implementation of Lyman-$\alpha$ coupling, X-ray heating, partial ionization, and the resulting 21-cm differential brightness temperature within \textsf{SCRIPT}. In \secn{sec:results}, we analyze the impact of Lyman-$\alpha$ and X-ray parameters on the 21-cm global signal and power spectrum, compare the predictions of the contrasting galaxy formation models, and discuss the breaking of astrophysical degeneracies. We summarize our conclusions in \secn{sec:summary}. In the Appendices, we demonstrate the numerical convergence of our results across different radiative transfer algorithms and spatial resolutions. Throughout this work, we adopt Planck best-fit cosmological parameters: $h = 0.67$, $\Omega_m = 0.310$, $\Omega_\Lambda = 0.690$, $\Omega_b = 0.049$, $n_s = 0.961$, $\sigma_8 = 0.829$, and $Y_{\text{He}} = 0.245$ \citep{Planck2018}.


\section{Methodology} \label{sec:method}

In this section, we describe the numerical methodology used to model the 21-cm signal during the Cosmic Dawn. Because the spin-temperature fluctuations are driven by the radiation fields of early luminous sources, we first outline the underlying galaxy population model that serves as the foundation for our radiative transfer calculations.

\subsection{Galaxy Properties}\label{sec:galaxy}

We compute the density, halo, and ionization fields within a cubical simulation volume of comoving length $L_\mathrm{box}$, following the methodology presented in \cite{Choudhury:2025}. Briefly, the large-scale density field is generated on a uniform grid using the second-order Lagrangian perturbation theory (2LPT) code MUSIC \cite{Hahn2011}, followed by cloud-in-cell (CIC) smoothing. For a simulation box partitioned into $N_\mathrm{grid}^3$ cells, the comoving spatial resolution is defined as:
\begin{equation}
\Delta r_c = \frac{L_\mathrm{box}}{N_\mathrm{grid}}.
\end{equation}
Because the relatively coarse resolution required for these large-volume simulations precludes the direct identification of dark matter halos, we populate each grid cell with halos using a conditional ellipsoidal collapse model \citep{Sheth2002}.

The intrinsic ultraviolet (UV) luminosity at a rest wavelength of $1500\,\text{\AA}$ for a halo of mass $M_h$ at redshift $z$ is modeled as:
\begin{align}
L_{\mathrm{UV}}(M_h, z) &= f_\star(M_h, z) \frac{\Omega_b}{\Omega_m}  l_\mathrm{UV}  M_h, \nonumber \\
&= \varepsilon_\star(M_h, z)  l_{\mathrm{UV,fid}}  M_h,
\end{align}
where $f_\star(M_h, z)$ is the star formation efficiency and $l_\mathrm{UV}$ is the specific UV luminosity. We adopt a fiducial specific luminosity of $l_{\mathrm{UV,fid}} = 8.66 \times 10^{19} \uni{erg}{}\uni{s}{-1}\uni{Hz}{-1}\uni{M_\odot}{-1}$. To capture the evolving galaxy population, we parameterize the effective star formation efficiency, $\varepsilon_\star(M_h, z)$, as a mass- and redshift-dependent power law:
\begin{equation}
\varepsilon_\star(M_h, z) = \varepsilon_{\star,10}(z)  \left(\frac{M_h}{M_{10}}\right)^{\beta_\star(z)},
\label{eq:eps_star}
\end{equation}
where $M_{10} \equiv 10^{10} M_\odot$. Both the normalization, $\varepsilon_{\star,10}(z)$, and the slope, $\beta_\star(z)$, are allowed to evolve smoothly with redshift:
\begin{align}
\log_{10} \varepsilon_{\star,10}(z) &= \ell_{\star,0} + \frac{\ell_{\star,\mathrm{jump}}}{2} \tanh\left(\frac{z - z_\mathrm{trans}}{\Delta z}\right),  \nonumber \\
\beta_\star(z) &= \beta_{\star,0} + \frac{\beta_{\star,\mathrm{jump}}}{2} \tanh\left(\frac{z - z_\mathrm{trans}}{\Delta z}\right).
\label{eq:eps_star_evol}
\end{align}
Here, $\ell_{\star, 0}$, $\ell_{\star, \rm jump}$, $\beta_{\star, 0}$, $\beta_{\star, \rm jump}$, $z_{\rm trans}$, and $\Delta z$ are free modeling parameters. This flexible parameterization is explicitly motivated by recent JWST observations, which require galaxies to form stars significantly more efficiently at $z \gtrsim 10$. For a comprehensive discussion of this parameterization, we refer the reader to \cite{Choudhury:2025}.

By combining the $L_\mathrm{UV} - M_h$ relation with the halo mass function, we compute the UVLF, $\Phi_\mathrm{UV}(M_\mathrm{UV})$, where the absolute UV magnitude is given by $M_\mathrm{UV} = 51.6 - 2.5 \, \log_{10} (L_\mathrm{UV} / \uni{erg}{}\uni{s}{-1}\uni{Hz}{-1})$. During this computation, we also account for the suppression of star formation inside ionized and photoheated regions due to radiative feedback \citep{Chakraborty2024}. While this feedback is largely negligible during the Cosmic Dawn when the IGM is overwhelmingly neutral, it becomes a crucial regulatory mechanism at lower redshifts.

Properly modeling this radiative feedback requires coupling the UVLF calculation to the broader reionization and thermal histories. To achieve this, we employ the extended version of the \textsf{SCRIPT} framework, which self-consistently models key late-reionization processes, including inhomogeneous recombinations, spatially fluctuating ionizing mean free paths, and the photoionization rate \citep{Choudhury:2025}. By incorporating a density-dependent subgrid clumping factor, thermal IGM evolution, and radiative feedback on low-mass galaxies, the framework accurately captures large-scale cosmic inhomogeneities. Ultimately, this well-calibrated, inhomogeneous distribution of star-forming galaxies acts as the source field for the Lyman-$\alpha$ and X-ray radiative transfer algorithms, which directly govern the spin-temperature fluctuations detailed in the subsequent sections.


\subsection{\boldmath Lyman-$\alpha$ Coupling} \label{sec:lyal}

With the star-formation properties of individual halos established, we now model the physical processes driving the spin temperature fluctuations. We begin by detailing the calculation of the Lyman-$\alpha$ coupling.

We compute the rate of photons emitted within the frequency range between the Lyman-$\alpha$ and Lyman-limit frequencies from a grid cell at location $\mathbf{x}$, per unit frequency, as:
\begin{equation}
\dot{N}_{\alpha,\nu}^{\mathrm{cell}}(z, \mathbf{x}) = l_{\alpha,\mathrm{fid}} \, \left(\Delta r_c \right)^3\int_{\mathrm{cell}} d\ln M_h \; 
\varepsilon_\star(M_h,z) \, f_L(M_h, z) \, M_h \frac{dn(z,\mathbf{x})}{d\ln M_h},
\label{eq:dot_N_alpha}
\end{equation}
where $dn(z,\mathbf{x})/d\ln M_h$ denotes the conditional halo mass function within the grid cell centered at $\mathbf{x}$, and $(\Delta r_c)^3$ is the comoving volume of the cell. The integration is performed over all halo masses capable of forming stars, which we define as halos with masses above the atomic cooling threshold. This total emission rate is in units of $\uni{photons}{}\uni{s}{-1}\uni{Hz}{-1}$.

Here, $f_L(M_h, z)$ denotes the Lyman-$\alpha$ efficiency parameter, which may, in general, depend on both halo mass and redshift. In this work, we parameterize this efficiency as a simple power law:
\begin{equation}
f_L(M_h, z) = f_{L,0} \left(\frac{M_h}{M_{10}}\right)^{\beta_L},
\end{equation}
with $f_{L,0}$ and $\beta_{L}$ serving as free model parameters. In this specific form, there is no explicit $z$-dependence for the Lyman-$\alpha$ efficiency itself.

The fiducial Lyman-$\alpha$ photon emissivity parameter, $l_{\alpha,\mathrm{fid}}$, is strictly proportional to the UV specific luminosity $l_{\mathrm{UV, fid}}$:
\begin{equation}
l_{\alpha,\mathrm{fid}} = \frac{l_{\mathrm{UV,fid}} \Delta\nu_\mathrm{ion}}{h_p \bar{\nu} \Delta\nu_\mathrm{tot}}.
\end{equation}
Here, the ionization bandwidth is $\nu_\mathrm{ion} = \nu_\mathrm{HeII} - \nu_H \approx 9.9 \pow{15} \uni{Hz}{}$, and the characteristic ionization frequency is $\bar{\nu} \approx 6.6 \pow{15}\uni{Hz}{}$. The Lyman-$\alpha$ band spans $\Delta\nu_\mathrm{tot} = \nu_\mathrm{limit} - \nu_\alpha = 8.2 \pow{14} \uni{Hz}{}$, where $\nu_\mathrm{limit} = 3.3 \pow{15} \uni{Hz}{}$ is the Lyman-limit frequency. Using our fiducial value of $l_{\mathrm{UV, fid}} = 8.66 \pow{19} \uni{erg}{}\uni{s}{-1}\uni{Hz}{-1}\uni{M_\odot}{-1}$, we obtain $l_{\alpha, \mathrm{fid}} = 2.40 \pow{31} \uni{photons}{}\uni{s}{-1}\uni{Hz}{-1}\uni{M_\odot}{-1}$. This parameter establishes a physically motivated baseline emissivity for standard stellar populations, derived directly from the UV luminosity calibration. Any departures from this baseline, arising from variations in the stellar initial mass function or metallicity, are absorbed into the dimensionless efficiency factor $f_L(M_h, z)$ in \eqn{eq:dot_N_alpha}.

We compute the number density of Lyman-series (Lyman-$m$) photons at sink cell $\mathbf{x}$ and redshift $z$:
\begin{equation}
n_m(z, \mathbf{x}) = \sum_{z_{\rm sr}>z}^{z_{\rm max}} \sum_{\mathbf{x}_{\rm sr}(z_{\rm sr})} \frac{\dot{N}_{\alpha,\nu_{\mathrm{em}}}^{\mathrm{cell}}(z_{\mathrm{sr}}, \mathbf{x}_{\mathrm{sr}}) \, \Delta \nu_{\mathrm{cell}}}{4\pi r_p^2 c},
\label{eq:n_m}
\end{equation}
where the numerator, $\dot{N}_{\alpha,\nu_{\mathrm{em}}}^{\mathrm{cell}} \Delta \nu_{\mathrm{cell}}$, represents the absolute rate of photons emitted by each source cell (located at $\mathbf{x}_\mathrm{sr}$ and redshift $z_\mathrm{sr}$) within the frequency interval $\Delta \nu_{\mathrm{cell}}$. Dividing this rate by the spherical surface area $4\pi r_p^2$ and the speed of light $c$ yields the local number density of these photons at the target cell (located at $\mathbf{x}$ and redshift $z$). Here, $r_p$ is the proper distance from a source cell, and we sum over all source cells at higher redshifts. Note that the Lyman-$m$ frequency is related to the emitted frequency by the appropriate cosmological redshift factor, $\nu_m = \nu_\mathrm{em} (1 + z) / (1 + z_\mathrm{sr})$, where the source redshift $z_\mathrm{sr}$ and the sink redshift $z$ are separated by the proper distance $r_p$. By definition, the resulting photon field emanating from a single source cell is spherically symmetric.

The photons redshifting into the Lyman-$m$ line frequency ($\nu_m$) in the sink cell were emitted from the source cell within a frequency interval corresponding to the source cell width. Hence:
\begin{equation}
\Delta\nu_{\mathrm{cell}}
\simeq
\nu_{\rm em} \frac{H(z_{\mathrm{sr}})}{c}
\frac{\Delta r_c}{1+z_{\mathrm{sr}}},
\end{equation}
where $z_{\mathrm{sr}}$ is the source redshift and, as defined earlier, $\Delta r_c$ is the comoving cell width.

At any given distance, multiple Lyman-$m$ lines cascade to contribute to the overall Lyman-$\alpha$ photon budget \cite{Pritchard:2006}. The maximum comoving distance traveled by photons just redward of the Lyman-$(m+1)$ line before they redshift into the Lyman-$m$ line is given by the shell horizon for each Lyman-$m$ shell:
\begin{equation}
r_\mathrm{shell}(m) = \frac{c}{H(z_{\mathrm{sr}})} (1+z_{\mathrm{sr}}) \frac{\Delta\nu_m}{\nu_m},
\end{equation}
where $\Delta\nu_m = \nu_{m+1} - \nu_m$. Note that shell 2 (comprising photons between the Lyman-$\alpha$ and Lyman-$\beta$ lines) contributes to the \textit{continuum} Lyman-$\alpha$ background, as these photons directly redshift into the Lyman-$\alpha$ line from the blueward side. In contrast, all higher shells ($m > 2$) contribute to \textit{injected} Lyman-$\alpha$ photons, which are generated precisely at the Lyman-$\alpha$ line center following atomic cascades from higher Lyman-series transitions \cite{CM04}. We compute the number density of continuum and injected photons separately in this work. However, we do not include the thermal effects caused by these photons \cite{RSS24, RS26}.

The total Lyman-$\alpha$ photon number density at $\mathbf{x}$ is given by:
\begin{equation}
n_\alpha(z, \mathbf{x}) = \sum_{m=2}^{30} n_{m}(z, \mathbf{x}) \, f_\mathrm{rec}(m) \, w_\mathrm{shell}(m).
\end{equation}
Here, $f_{\mathrm{rec}}$ is the probability of a Lyman-$m$ photon cascading down to the Lyman-$\alpha$ line. We utilize the values provided in \cite{Pritchard:2006}, taking $f_{\mathrm{rec}}(2)=1$ for continuum photons.

Because our coarse-grid simulations can employ large cell widths, the Lyman-$m$ line horizon might cover only a fraction of a cell. To correct for this sub-grid effect, we apply a spatial weighting factor:
\begin{equation}
w_\mathrm{shell}(m) = \frac{\min(r_\mathrm{shell}(m) - r_\mathrm{dist}, \, \Delta r_c)}{\Delta r_c}, \quad r_{\rm shell}>r_{\rm dist}
\label{eq:w_shell}
\end{equation}
where $r_\mathrm{dist}$ is the comoving distance from the source to the inner edge of the target cell. When $r_{\rm dist} > r_{\rm shell}$, we take $w_{\rm shell}(m) = 0$. Neglecting this correction would lead to an unphysical over-counting of injected photons relative to continuum photons.

Finally, the scattering rate of Lyman-$\alpha$ photons within a sink cell is:
\begin{equation}
P_\alpha(z, \mathbf{x}) = c \,\sigma_\alpha \, n_\alpha(z, \mathbf{x}),
\end{equation}
where $\sigma_\alpha$ is the effective Lyman-$\alpha$ scattering cross-section, which includes the line profile correction. The corresponding Lyman-$\alpha$ coefficient for Wouthuysen-Field (WF) coupling \citep{Wouthuysen1952, Field1958} is then:
\begin{equation}
y_\alpha(z, \mathbf{x})
=
\frac{4 P_\alpha(z, \mathbf{x}) \, h_p\nu_{21}}
{27 A_{21} k_B T_K(z, \mathbf{x})}
=
\frac{4 c \,\sigma_\alpha \, n_\alpha(z, \mathbf{x}) \, h_p\nu_{21}}
{27 A_{21} k_B T_K(z, \mathbf{x})},
\end{equation}
where $\nu_{21}$ and $A_{21}$ are the frequency and the Einstein coefficient for the 21-cm transition, respectively, and $T_K(z, \mathbf{x})$ is the kinetic temperature of the gas. We detail the calculation of $T_K(z, \mathbf{x})$, which is primarily driven by X-ray heating from early galactic sources, in the following section.


\subsection{X-ray Heating and Partial Ionization} \label{sec:xrays}

The X-ray photons emitted by early galactic sources contribute primarily to the heating and partial ionization of the IGM. To model this effect, we assume that these sources emit isotropically. The comoving X-ray photon emissivity density at a location $\mathbf{x}$ and redshift $z$ is:
\begin{equation}
\epsilon_X(\nu, z, \mathbf{x}) =
\dot{\rho}_{\star,X}(z, \mathbf{x})\,
\mathcal{L}_X(\nu),
\label{eq:xr_emissivity_1}
\end{equation}
where $\dot{\rho}_{\star,X} = d \rho_{\star,X} / dt$ is the star-formation rate density weighted by the X-ray efficiency, with units of $\uni{M_\odot}{}\uni{yr}{-1}\uni{cMpc}{-3}$. We first compute the effective stellar mass density from the halo mass function in each cell as:
\begin{equation}
\rho_{\star,X}(z, \mathbf{x}) =
\int_\mathrm{cell} d\ln M_h \,
\varepsilon_\star(M_h,z) \, f_X(M_h,z) \, M_h \, \frac{dn(z, \mathbf{x})}{d\ln M_h},
\end{equation}
which has units of $\uni{M_\odot}{}\uni{cMpc}{-3}$, and then differentiate with respect to $t$. Here, $f_X(M_h,z)$ is the X-ray efficiency parameter. Similar to $f_L$, we parameterize it as:
\begin{equation}
f_X(M_h,z) = f_{X, 0} \left(\frac{M_h}{M_{10}}\right)^{\beta_X},
\end{equation}
where both $f_{X,0}$ and $\beta_X$ are free modeling parameters.

We take the spectral energy distribution of the X-ray photons to follow a power law:
\begin{equation}
\mathcal{L}_X(\nu) = \frac{L_{X,0}}{h_p \nu} \, A_X \,\nu^{-\alpha}, 
\label{eq:L_x}
\end{equation}
where the spectral index $\alpha$ is a free parameter, and $L_{X,0}=10^{40.5}\,\mathrm{erg\,s^{-1}}(\mathrm{M_\odot\,yr^{-1}})^{-1}$. Here, $L_{X,0}$ represents the fiducial specific X-ray luminosity per unit star formation rate, derived from local calibrations of high-mass X-ray binary populations \citep{Mineo2012}. Analogous to $l_{\alpha,\mathrm{fid}}$ in \secn{sec:lyal}, $L_{X,0}$ anchors the X-ray emissivity to a physical baseline. Any astrophysical modifications, such as low-metallicity effects or top-heavy initial mass functions, are captured by the dimensionless efficiency factor $f_X(M_h, z)$, which scales the X-ray output relative to this fiducial benchmark. The factor $(h_p\nu)^{-1}$ converts the energy spectrum into a photon-number spectrum, such that $\mathcal{L}_X$ has units of $\uni{photons}{}\uni{s}{-1}\uni{Hz}{-1}(\uni{M_\odot}{}\uni{yr}{-1})^{-1}$. The normalization constant $A_X$ ensures that:
\begin{equation}
\int_{\nu_{\mathrm{min}}}^{\nu_{\mathrm{max}}} A_X \nu^{-\alpha} d\nu = 1.
\label{eq:A_X_norm}
\end{equation}

From \eqns{eq:xr_emissivity_1}{eq:L_x}, the emissivity of X-ray photons per unit time and frequency range becomes:
\begin{equation}
\epsilon_X(\nu, z, \mathbf{x})= \frac{L_{X,0}}{h_p\nu} \, A_X \, \nu^{-\alpha} \, \dot{\rho}_{\star,X}(z, \mathbf{x}).
\label{eq:xr_emissivity_2}
\end{equation}
This emissivity has units of $\uni{photons}{}\uni{s}{-1}\uni{Hz}{-1}\uni{cMpc}{-3}$. In practice, evaluating the continuous frequency integral in \eqn{eq:A_X_norm} over the X-ray band $[500 \text{ eV}, 2 \text{ keV}]$ requires discretizing the photon frequency spectrum. We adopt a logarithmic binning scheme in frequency ($\Delta \ln \nu$), which efficiently captures the steep power-law spectrum $\nu^{-\alpha}$ with minimal computational overhead. We use 10 logarithmic bins over this frequency range and have confirmed that our results remain convergent for a larger number of bins.

Next, we compute the total specific photon flux in units of $\uni{photons}{}\uni{s}{-1}\uni{Hz}{-1}\uni{cm}{-2}$ at the sink position $\mathbf{x}$ and redshift $z$ as \citep{Pritchard2007}:
\begin{equation}
    J_X(\nu,z,\mathbf{x}) =
    (\Delta r_c)^3
    \sum_{z_{\rm sr}>z}^{z_{\rm max}} \,
    \sum_{\mathbf{x}_{\rm sr} (z_{\rm sr})} \,
    \frac{1}{4\pi r_p^2}\,
    \epsilon_X(\nu_{\mathrm{em}}, z_{\mathrm{sr}}, \mathbf{x}_{\mathrm{sr}}) \,
    e^{-\tau(\nu,z,z_{\mathrm{sr}})}, 
    \label{eq:J_x}
\end{equation}
where $\mathbf{x}_{\mathrm{sr}}$ is the source position and $r_p$, as defined earlier, is the proper distance between the source and sink. Similar to \eqn{eq:dot_N_alpha}, $(\Delta r_c)^3$ converts volume emissivity $\epsilon_X(\nu_{\mathrm{em}}, z_{\mathrm{sr}}, \mathbf{x}_{\mathrm{sr}})$ to photon emission rate per unit frequency from a source cell. Photons emitted at a source redshift $z_{\mathrm{sr}}$ with frequency $\nu_{\mathrm{em}}$ are observed at redshift $z$ at a frequency $\nu = \nu_{\mathrm{em}}(1+z)/(1+z_{\mathrm{sr}})$. The sum includes contributions from source cells at all relevant source redshifts $z_{\mathrm{sr}}$ up to $z_{\max}$, the highest redshift in our simulation when the sources begin forming (typically $z_\mathrm{max} = 25$ in our calculations). We allow these photons to propagate up to the end of reionization ($z_{\mathrm{min}} = 5$), where we terminate our calculations. However, we only explicitly compute the fluctuating radiation field up to half the box width ($L_\mathrm{box} / 2$); beyond this distance, we uniformly distribute the photons as a homogeneous X-ray background.

The optical depth of a photon at frequency $\nu$ traveling between $z_1$ and $z_2$, accounting for photon absorption by neutral hydrogen and helium atoms, is:
\begin{equation}
\tau(\nu,z_1,z_2) =
\int_{z_1}^{z_2} dz \;
\frac{c}{H(z)(1+z)}
\left[
n_{\mathrm{HI}}(z) \, \sigma_{\mathrm{HI}}(\nu')
+
n_{\mathrm{HeI}}(z) \, \sigma_{\mathrm{HeI}}(\nu')
\right],
\end{equation}
where $\nu'=\nu(1+z)/(1+z_1)$. The photo-ionization cross-sections for hydrogen and helium decrease with increasing frequency as:
\begin{align}
\sigma_{\mathrm{HI}}(\nu) &= \sigma_{\mathrm{HI},0}\left(\frac{\nu_{\mathrm{HI}}}{\nu}\right)^3,\quad \nu>\nu_{\mathrm{HI}}, \nonumber \\
\sigma_{\mathrm{HeI}}(\nu) &= \sigma_{\mathrm{HeI},0}\left(\frac{\nu_{\mathrm{HeI}}}{\nu}\right)^3,\quad \nu>\nu_{\mathrm{HeI}}.
\end{align}
The local densities of neutral hydrogen and helium, $n_{\mathrm{HI}}, n_{\mathrm{HeI}} \propto (1+\delta) (1-x_e)$, account for the local overdensity as well as the partial ionization of the medium ($x_e$).

For our coarse-resolution boxes, the optical depth of a cell can be very large at low frequencies ($\tau_{\mathrm{cell}} (\nu) > 1$). Therefore, we accurately calculate the fraction of the X-ray flux $J_X(\nu,z,\mathbf{x})$ absorbed by hydrogen and helium atoms as, respectively:
\begin{align}
a_{\mathrm{HI}} (\nu) &=
\frac{\sigma_{\mathrm{HI}}(\nu) n_{\mathrm{HI}}}{\sigma_{\mathrm{HI}}(\nu) n_{\mathrm{HI}} +\sigma_{\mathrm{HeI}}(\nu) n_{\mathrm{HeI}}}\, (1-e^{-\tau_{\mathrm{cell}} (\nu)}),
\quad
\mathrm{and} \nn \\
a_{\mathrm{HeI}} (\nu) &=
\frac{\sigma_{\mathrm{HeI}}(\nu)n_{\mathrm{HeI}}}{\sigma_{\mathrm{HI}}(\nu)n_{\mathrm{HI}} +\sigma_{\mathrm{HeI}}(\nu)n_{\mathrm{HeI}}}\, (1-e^{-\tau_{\mathrm{cell}}(\nu)}).
\end{align}

The excess energy of the photoelectrons emitted by hydrogen and helium atoms is $E_{\mathrm{HI}} = h_p(\nu - \nu_{\mathrm{HI}})$ and $E_{\mathrm{HeI}} = h_p(\nu - \nu_{\mathrm{HeI}})$. These energetic electrons cause heating and secondary ionizations within the neutral IGM \citep{1985ApJ...298..268S, Heating2001}. The local X-ray heating rate within a cell can be written as the integration of the photon flux, \eqn{eq:J_x}, over the frequency range, divided by the cell width $\Delta r_p = \Delta r_c /(1+z)$, and weighted by the probability of these photons being absorbed:
\begin{equation}
\mathcal{H}_X(z, \mathbf{x})
=
\frac{1}{\Delta r_p}
\int_{\nu'_{\mathrm{min}}}^{\nu'_{\mathrm{max}}} d\nu \;
J_X(\nu, z, \mathbf{x})
\left[
a_{\mathrm{HI}} \,
f_{\mathrm{heat}}^{(\mathrm{HI})}(E_{\mathrm{HI}},x_e) \,
E_{\mathrm{HI}}
+
a_{\mathrm{HeI}} \,
f_{\mathrm{heat}}^{(\mathrm{HeI})}(E_{\mathrm{HeI}},x_e) \,
E_{\mathrm{HeI}}
\right].
\end{equation}
Here, $f_{\mathrm{heat}}$ is the fraction of the primary electron energy that is deposited as heat. With this convention, $\mathcal{H}_X$ is a volumetric heating rate with units of $\uni{erg}{}\uni{s}{-1}\uni{cm}{-3}$. While $\nu'_{\mathrm{max}} = \nu_{\mathrm{max}}$, the minimum frequency is $\nu'_{\mathrm{min}} = \nu_{\mathrm{min}}(1+z)/(1+z_{\mathrm{max}})$, properly accounting for the redshift.

Similarly, the X-ray ionization rate of hydrogen is:
\begin{align}
\Gamma_X^{\mathrm{ion}}(z, \mathbf{x})
&=
\frac{1}{\Delta r_p}
\int_{\nu'_{\mathrm{min}}}^{\nu'_{\mathrm{max}}} d\nu \;
J_X(\nu,z,\mathbf{x}) \,
\left[
a_{\mathrm{HI}}(\nu) \,
\left(1 + N_{\mathrm{ion,HI}}^{(\mathrm{HI})}(E_{\mathrm{HI}},x_e)
\right) \right. \nn \\
& \qquad \qquad + \left.
a_{\mathrm{HeI}}(\nu) \,
N_{\mathrm{ion,HI}}^{(\mathrm{HeI})}(E_{\mathrm{HeI}},x_e)
\right],
\end{align}
where the first term corresponds to the primary $\mathrm{HI}$ ionization due to the X-ray photons, and the $N_{\mathrm{ion,HI}}$ terms account for secondary $\mathrm{HI}$ ionizations produced by the photoelectrons from $\mathrm{HI}$ and $\mathrm{HeI}$ absorption, respectively. Both $f_{\mathrm{heat}}$ and $N_{\mathrm{ion,HI}}$ depend on the ionized fraction of the cell and the electron energy. We use the tables provided in \cite{Furlanetto:2010} to interpolate these quantities. We note that these photoelectrons also produce Lyman-$\alpha$ photons, but we have not included that effect in our calculations, as it is sub-dominant.

Having obtained the ionization rate in each grid cell, we use it to compute the evolution of the X-ray ionized electron fraction, accounting for recombinations. The total electron fraction in a given cell is the sum of those ionized by X-rays and those ionized by standard UV photons, as computed by the reionization algorithm described in \secn{sec:galaxy}. We use this total ionization fraction of the cell to calculate the optical depth of the IGM and the resulting 21-cm signal. Physically, this means that the mean free path of X-ray photons increases as the Universe becomes highly ionized, eventually allowing these photons to free-stream through the medium at the end of reionization.

Finally, the kinetic temperature of the gas evolves according to:
\begin{equation}
\frac{3}{2} k_B \, n_b \, \frac{dT_K(z, \mathbf{x})}{dt}
=
\mathcal{H}_X(z, \mathbf{x})
-
3 k_B \, n_b \, T_K(z, \mathbf{x}) \, H(z),
\end{equation}
where the second term accounts for adiabatic cooling due to Hubble expansion ($n_b$ being the local baryon number density). 

\subsection{Numerical Implementation and the 21-cm Signal}

To evaluate the Lyman-$\alpha$ and X-ray background fields (defined in \eqns{eq:n_m}{eq:J_x}, respectively) across our discrete simulation grid, we implement two complementary numerical algorithms designed to ensure strict photon conservation and resolution convergence:

\begin{enumerate}

\item \Msink: In this approach, we iterate over the sink grid cells and integrate the incoming photon flux by looking up contributions from all relevant source cells at higher redshifts. This conceptual framework is analogous to the excursion-set filtering techniques traditionally employed by widely used semi-numerical codes (e.g., \textsf{21CMFAST} \cite{21CMFAST}).

\item \Msrc: In this complementary approach, we iterate over the source grid cells and distribute their emitted photons forward in time to the appropriate sink cells at lower redshifts. This methodology closely aligns with the 3D profile-painting and radiative propagation techniques utilized in more recent frameworks (e.g., \textsf{GRIZZLY} \cite{Ghara2018} and \textsf{BEORN} \cite{BEORN}).

\end{enumerate}

We defer a detailed technical description and numerical comparison of these two methods to \appndx{app:source_sink}. Notably, both approaches yield identical, converged results for the radiation backgrounds in our calculations.

With the radiation fields established, we use the X-ray heating and Lyman-$\alpha$ coupling grids to compute the local spin temperature:
\begin{equation}
T_S(z, \mathbf{x})=\frac{T_{\mathrm{CMB}}(z) + y_\alpha(z, \mathbf{x}) \, T_K(z, \mathbf{x})}{1+y_\alpha(z, \mathbf{x})}.
\end{equation}
Finally, the observable 21-cm differential brightness temperature signal is given by:
\begin{equation}
\delta T_b(z, \mathbf{x}) \approx 27 \, x_{\mathrm{HI}}(z, \mathbf{x}) \left[1+\delta(z, \mathbf{x})\right]\left(1 - \frac{T_{\mathrm{CMB}}(z)}{T_S(z, \mathbf{x})}\right)\left(\frac{1+z}{10} \right)^{1/2}\left(\frac{0.15}{\Omega_m h^2}\right)^{1/2}\left(\frac{\Omega_b h^2}{0.023}\right) \; \mathrm{mK},
\end{equation}
where $x_{\mathrm{HI}}$ is the neutral hydrogen fraction and $\delta$ is the local baryon density contrast.

We utilize the simulated density, total ionization (accounting for both UV reionization and partial X-ray ionization), and spin temperature fields to compute this observable signal. Note that we neglect collisional coupling in this work, as its effect on the spin temperature is negligible at the redshifts of our interest. Furthermore, we have omitted redshift-space distortions induced by peculiar velocity fluctuations in the current paper; however, these kinematic effects can be straightforwardly incorporated into the formalism in future analyses.

\section{Results} \label{sec:results}

In this section, we present the results for the 21-cm global signal and its power spectrum during the Cosmic Dawn. We primarily focus on how the inclusion of Lyman-$\alpha$ coupling and X-ray heating fluctuations within \textsf{SCRIPT} affects the observable signal.

The results in this paper have been computed for a simulation box of length $L_\mathrm{box} = 256 \, h^{-1} \, \mathrm{cMpc}$. The default grid resolution has been chosen corresponding to $N_\mathrm{grid} = 16$, however, we test the convergence of our results with respect to resolution (i.e., varying $N_\mathrm{grid}$) in \appndx{app:res}.

\subsection{Calibrating Source Models to High-Redshift UV Luminosity Functions}\label{sec:galaxy_results}

\begin{figure}[tbp]
    \centering
    \includegraphics[width=0.7\linewidth]{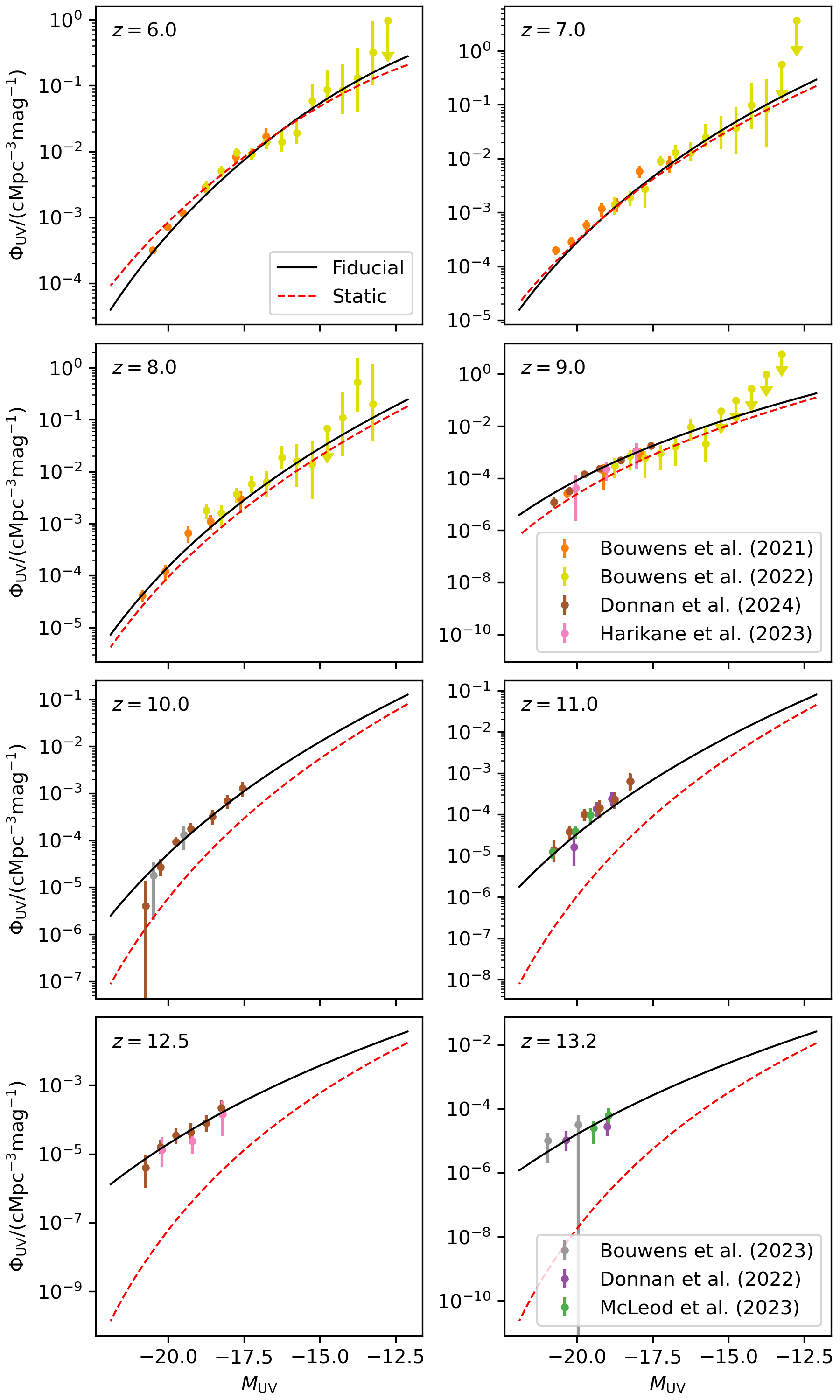}
    \caption{
        The UVLFs for the \Mfid model (solid black line) and the \Mstat model (dashed red line). The points with error bars represent observed UVLFs from \citep{Bouwens2021,Bouwens2022,Donnan2023,Harikane2023,Bouwens2023,McLeod2024,Donnan2024}. The \Mfid model matches observations across the full redshift range, while the \Mstat model deviates significantly at $z > 9$.
    }
    \label{fig:uvlfs}
\end{figure}

Our galaxy population model assumes a star formation efficiency ($\varepsilon_*$) that varies with redshift and halo mass, as in \eqns{eq:eps_star}{eq:eps_star_evol}. We compare two distinct models for the galaxy population. In the \Mfid model, we allow $\varepsilon_{\star,10}$ and $\beta_\star$ to evolve with redshift. The values characterizing the evolution, i.e., $\ell_{\star, 0}$, $\ell_{\star, \mathrm{jump}}$, $\beta_{\star, 0}$, $\beta_{\star, \mathrm{jump}}$, $z_\mathrm{trans}$ and $\Delta z$, are chosen to ensure the resulting UVLF matches observations over the wide range $6 \lesssim z \lesssim 13.2$. In the second model, which we call \Mstat, the quantities $\varepsilon_{\star,10}$ and $\beta_\star$ remain redshift-independent, which is ensured by using $\ell_{\star, \mathrm{jump}} = \beta_{\star, \mathrm{jump}} = 0$, and tuning $\ell_{\star, 0}$ and $\beta_{\star, 0}$ to match the UVLF only at $z \sim 6$. The parameter values for these two models are summarized in \tab{table:params}.

\begin{table}[tbp] 
    \centering
    \begin{tabular}{|c|c|c|c|}
        \hline
        parameter                 & \Mfid & \Mstat & \Mtuned \\
        \hline
        $\ell_{\star,0}$          & $-0.69$  & $-2.90$ & $-2.90$ \\
        $\ell_{\star, \rm jump}$  & 5.06     & 0       & 0 \\
        $\beta_{\star,0}$         & 1.82     & 0.61    & 0.61     \\
        $\beta_{\star, \rm jump}$ & 3.04     & 0       & 0     \\
        $z_{\rm trans}$           & 16.22    & --      & --     \\
        $\Delta z$                & 7.23     & --      & --     \\
        \hline 
        $f_L$                     & 1        & 1       & 10    \\
        $\beta_L$                 & 0        & 0       & 0     \\
        $f_X$                     & 10       & 10      & 0.0005 \\
        $\beta_X$                 & 0        & 0       & -3.0 \\
        $\alpha$                  & 1.5      & 1.5     & 1.5  \\
        \hline
    \end{tabular}
    \caption{Parameter values for the fiducial and redshift-independent galaxy population models.}
    \label{table:params}
\end{table}

The comparison of the UVLFs for these two models with the observations is shown in \fig{fig:uvlfs}. The comparison illustrates that the \Mfid model accurately reproduces the observed galaxy population across cosmic time, whereas the \Mstat model fails to capture the abundance of galaxies at high redshifts $z \gtrsim 9$. In particular, it predicts a much lower abundance of galaxies than observed, which confirms that redshift evolution in star formation efficiency is essential to reconcile models with recent JWST observations.

We also tested the ionization histories and the IGM properties at $z \sim 6$ for these two models against observations. To calculate the ionization history, we assume an escape ionizing efficiency $\varepsilon_\mathrm{esc}$ having the following power-law halo-mass dependence,
\begin{equation}
    \varepsilon_\mathrm{esc} = \varepsilon_{\mathrm{esc},10} \left(\frac{M_h}{10^{10}M_\odot}\right)^{\beta_\mathrm{esc}}.
\end{equation}
Note that $\varepsilon_\mathrm{esc}$ is a combination of the escape fraction $f_\mathrm{esc}$ and $\xi_\mathrm{ion}$, the rate of ionizing photons per unit UV luminosity. We assume it to be independent of redshift. We find that $\beta_\mathrm{esc} = -0.18$ for both models, and $\varepsilon_{\mathrm{esc},10} = 0.91$ for the \Mfid model and $\varepsilon_{\mathrm{esc},10} = 1.38$ for the \Mstat model provide a reasonable match to the observed ionization history\footnote{While $\varepsilon_{\mathrm{esc},10}$ values exceeding unity may seem high for an escape fraction, this is an effective efficiency parameter. It accounts for uncertainties in the ionizing photon production rate ($\xi_\mathrm{ion}$); a value $> 1$ suggests a more ionizing-efficient stellar population, as expected in low-metallicity environments at high redshift.}. Upon tuning of $ \varepsilon_\mathrm{esc}$, both the models produce a CMB electron optical depth of $\approx 0.052$, consistent with the latest measurements \cite{Planck2018}. The end of reionization is around $z \sim 5.7$ for the \Mfid model, while it is slightly earlier, at $z \sim 6$, for the \Mstat model. We also checked and found that it is possible to tune the parameters related to the reionization temperature and subgrid density distribution for both the models independently so as to ensure both are consistent with the observations related to the Lyman-$\alpha$ forest absorption spectra, e.g., the thermal state of the IGM at $z \sim 5.5-6.0$ \cite{Gaikwad2020}, the mean free path of ionizing photons \cite{Zhu2023}, the photoionization rate of HI \cite{Wyithe2011,Calverley2011,DAloisio2018,Gaikwad2023}.

\subsection{Parameter Space Exploration of the Global Signal and Spatial Fluctuations}\label{sec:laxr_results}

\begin{figure}[tbp]
    \centering
    \includegraphics[width=1\linewidth]{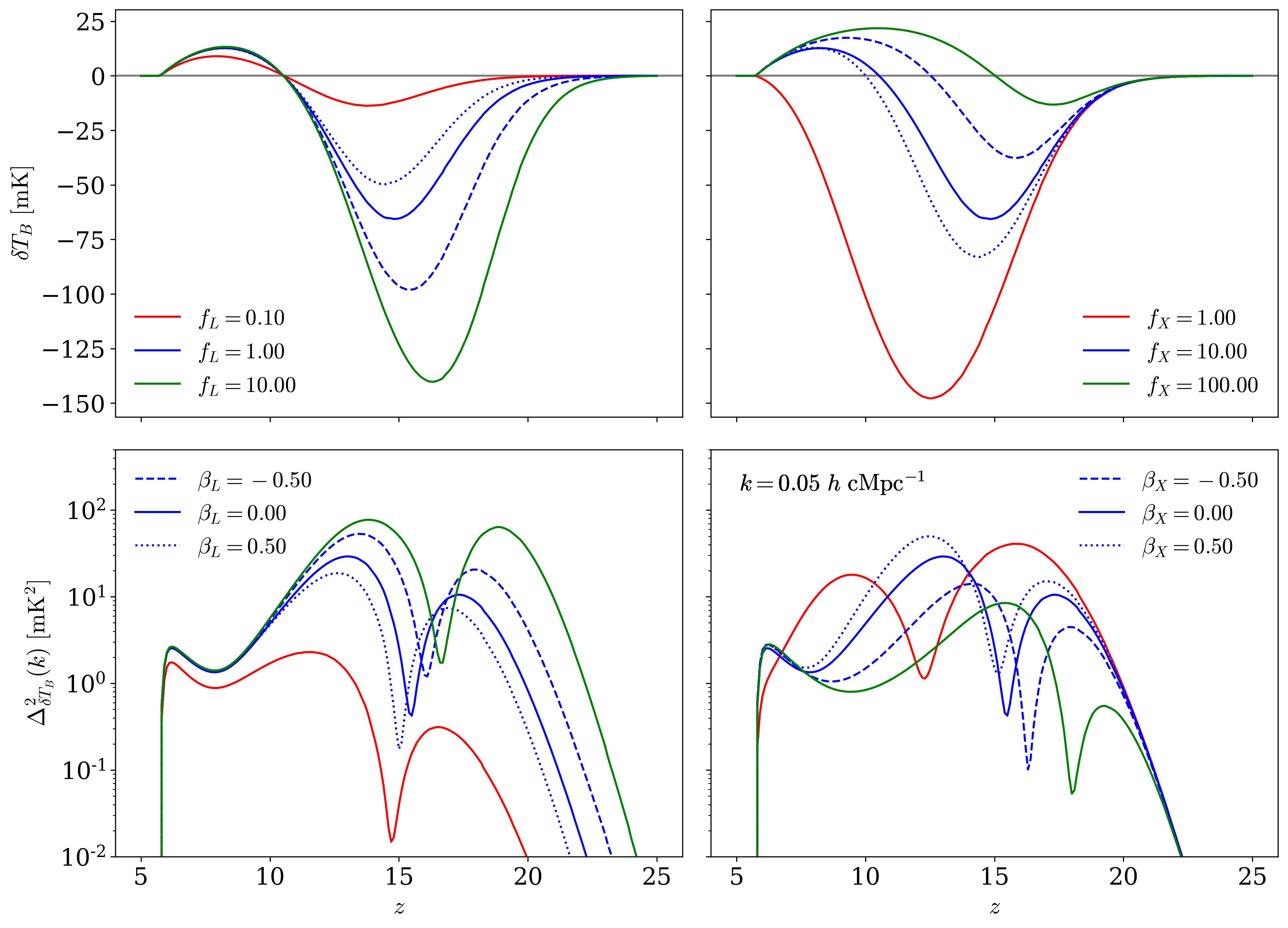}
    \caption{The global 21-cm differential brightness temperature $\delta T_b$ (top panels) and its power spectrum $\Delta^2_{\delta T_b}(k)$ evaluated at $k = 0.05 \, h \text{ cMpc}^{-1}$ (bottom panels) as a function of redshift $z$. The left panels illustrate the effect of varying the Lyman-$\alpha$ efficiency parameters: the overall normalization $f_L$ (solid curves) and the mass-scaling index $\beta_L$ (dashed and dotted curves), while the X-ray parameters are fixed at $f_X = 10$ and $\beta_X = 0$. The right panels demonstrate the impact of varying the X-ray efficiency parameters: the overall normalization $f_X$ (solid curves) and the mass-scaling index $\beta_X$ (dashed and dotted curves), while keeping the Lyman-$\alpha$ parameters fixed at $f_L = 1$ and $\beta_L = 0$. The solid blue curve in all panels corresponds to our fiducial parameter set ($f_L=1$, $\beta_L=0$, $f_X=10$, $\beta_X=0$). All theoretical predictions are computed utilizing the \Mfid galaxy formation model via the \Msink for the radiative transfer, with a grid resolution corresponding to $N_\mathrm{grid}=16$ and a redshift step size of $\Delta z = 0.1$. 
   } 
    \label{fig:compare_params}
\end{figure}

Let us first test whether our formalism produces results along the expected lines when we vary the Lyman-$\alpha$ and X-ray modeling parameters. For the galaxy formation, we choose the \Mfid model in this section.

In \fig{fig:compare_params}, we present the global 21-cm signal and its corresponding power spectrum (at $k=0.05\,h \text{ cMpc}^{-1}$) computed by \textsf{SCRIPT} across a wide range of Lyman-$\alpha$ and X-ray parameter space. To establish a baseline, we first adopt a set of fiducial values, $f_L=1$, $\beta_L=0$, $f_X=10$, and $\beta_X=0$, that produce a typical 21-cm signal consistent with standard theoretical expectations (shown by blue solid curves). We have taken $\alpha = 1.5$ throughout this work. This fiducial model yields a pronounced absorption trough reaching $\delta T_b \approx -65$~mK at $z \approx 15$, followed by an emission feature at $z \approx 9$. Its power spectrum displays the characteristic three-peak structure corresponding to the consecutive epochs of Lyman-$\alpha$ coupling, X-ray heating, and reionization. To understand the sensitivity of the 21-cm signal to high-redshift astrophysics, we systematically vary the modeling parameters around this fiducial baseline.

In the left panels of \fig{fig:compare_params}, we explore variations in the Lyman-$\alpha$ efficiency parameters while keeping the X-ray parameters fixed. Specifically, we vary the overall normalization $f_L$ from 0.1 to 10, and the power-law index $\beta_L$ from $-0.5$ to $0.5$. The top-left panel illustrates the global signal, where the variation in $f_L$ spans a wide range of possible WF coupling histories. For $f_L=10$ (solid green curve), the coupling is strong and early, rapidly driving the spin temperature toward the kinetic temperature of the gas, which shifts the absorption feature to a higher redshift with a larger negative amplitude. Conversely, for $f_L=0.1$ (solid red curve), the coupling is weak and late, shifting the feature to lower redshifts with a reduced amplitude. The parameter $\beta_L$ controls how the Lyman-$\alpha$ emission scales with halo mass. A negative index ($\beta_L = -0.5$, dashed blue curve) shifts the bulk of the emission to the highly abundant, low-mass halos, leading to a strong and early absorption trough. In contrast, a positive index ($\beta_L = 0.5$, dotted blue curve) assigns greater weight to high-mass halos, resulting in a weak and late trough.

The bottom-left panel displays the corresponding power spectrum. As expected, since $\Delta^2_{\delta T_b}$ scales with the global signal amplitude, increasing (decreasing) $f_L$ leads to stronger (weaker) peaks occurring at earlier (later) redshifts. Interpreting the dependence on $\beta_L$ requires understanding a subtle interplay between the global signal and spatial clustering. The amplitude of the 21-cm power spectrum is roughly proportional to the square of the mean global signal multiplied by the intrinsic spatial fluctuations of the radiation field. When low-mass halos dominate ($\beta_L = -0.5$), their vast numbers boost the global Lyman-$\alpha$ photon budget. This enhances the overall mean 21-cm signal amplitude and shifts the WF coupling peak to earlier times. However, because low-mass halos are significantly less clustered (i.e., they have a lower spatial bias) than high-mass halos, the intrinsic spatial fluctuations of the Lyman-$\alpha$ background decrease. Thus, the observed power spectrum is a non-trivial combination of these two competing effects: a boosted global signal amplitude attempting to raise the power, and suppressed spatial clustering attempting to lower it.

In the right panels of \fig{fig:compare_params}, we perform a similar exploration for the X-ray efficiency parameters, $f_X$ and $\beta_X$, while maintaining the fiducial Lyman-$\alpha$ parameters. The top-right panel shows the global signal, utilizing values ranging from $f_X=1$ to $f_X=100$ to capture a broad spectrum of heating scenarios. For $f_X = 1$ (solid red curve), the heating is so weak that the universe remains cold until the end of reionization, producing no emission feature. On the other hand, for $f_X = 100$ (solid green curve), the heating is early and intense, driving the IGM temperature above the CMB well into the Cosmic Dawn ($z \sim 20$). The effect of $\beta_X$ on the global signal follows a similar logic: a negative (positive) $\beta_X$, shown by the blue dashed (dotted) curve, leads to X-ray production dominated by low- (high-)mass halos. This results in an earlier (later) onset of IGM heating, which suppresses (enhances) the depth of the main 21-cm absorption trough.

Finally, the bottom-right panel highlights the effect of the X-ray efficiency parameters on the power spectrum. As discussed in the context of the Lyman-$\alpha$ efficiency, a lower (higher) $f_X$ leads to a later (earlier) onset of heating, thereby making the amplitude of the peaks stronger (weaker). In fact, for $f_X = 1$ (solid red curve), the heating is so inefficient that the corresponding peak is entirely absent. Interpreting the effect of $\beta_X$ on the power spectrum involves the same subtleties as the Lyman-$\alpha$ case: for a negative $\beta_X$, the widespread heating by ubiquitous low-mass halos shifts the onset of heating to higher redshifts, reducing the mean global signal and, consequently, the amplitude of the power spectrum. Furthermore, the relatively uniform spatial distribution (low clustering bias) of these low-mass sources smooths out the local kinetic temperature fluctuations. Unlike the Lyman-$\alpha$ case, these two effects drive the power spectrum in the same direction, resulting in unambiguously lower peaks for a negative $\beta_X$. The opposite holds true for a positive $\beta_X$.

\subsection{Distinguishing Source Models using 21-cm Observables}\label{sec:compare}

\begin{figure}[tbp]
    \centering
    \includegraphics[width=0.49\linewidth]{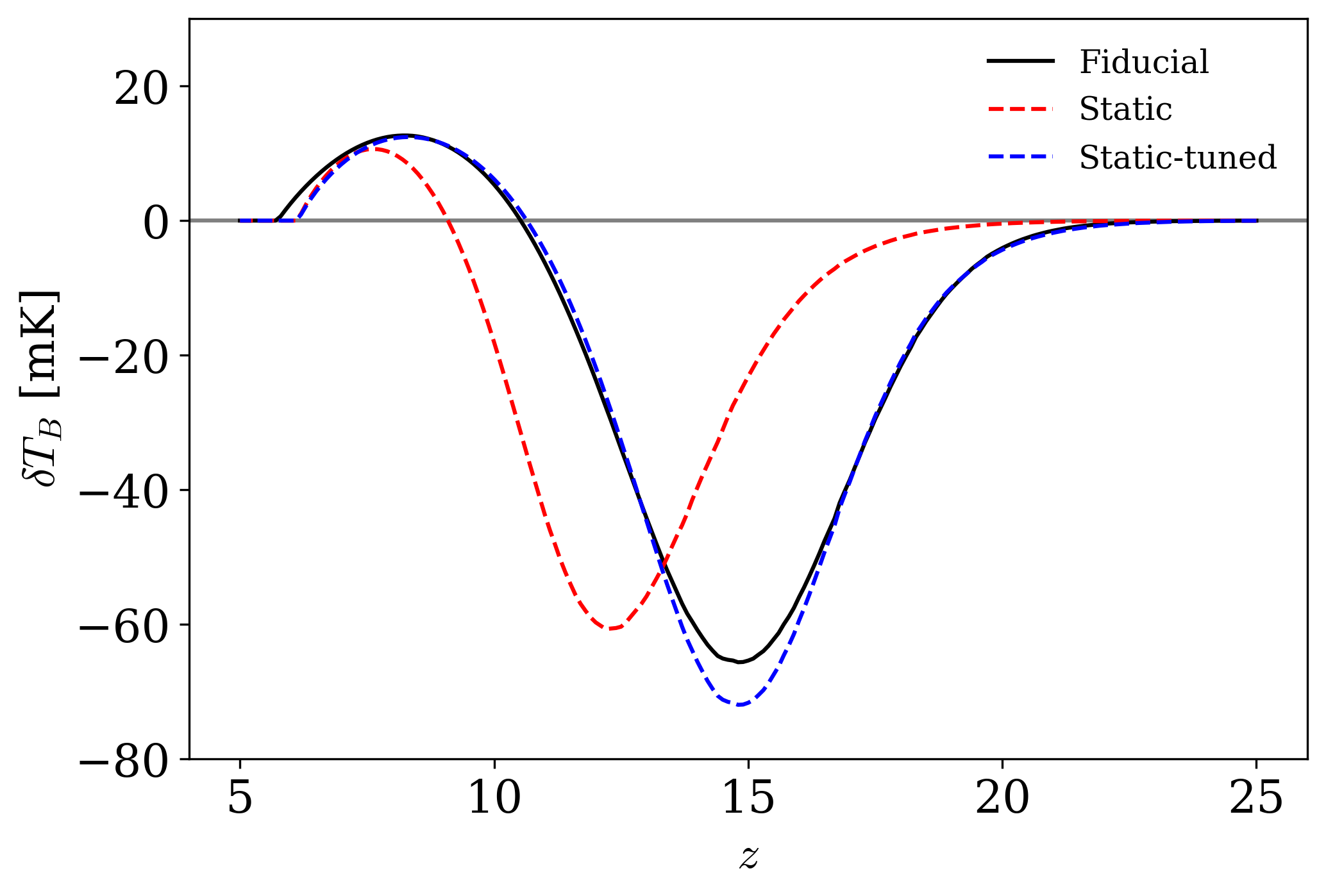}
    \includegraphics[width=0.49\linewidth]{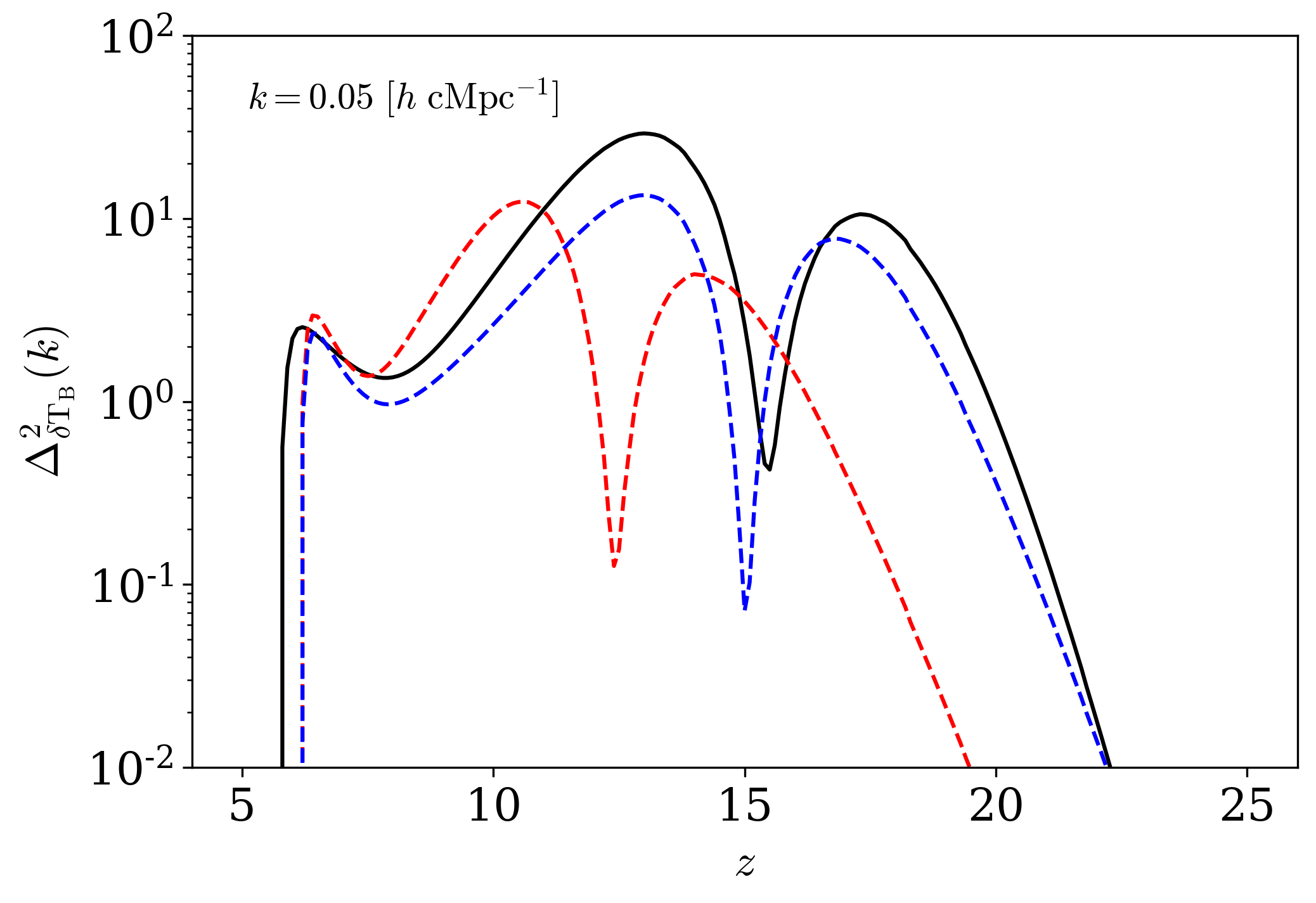}
    \caption{The global 21-cm differential brightness temperature $\delta T_b$ (left panel) and its power spectrum $\Delta^2_{\delta T_b}(k)$ evaluated at $k = 0.05 \, h \text{ cMpc}^{-1}$ (right panel) as a function of redshift $z$. We compare the predictions of the \Mfid galaxy formation model (solid black curve) against two variations of the \Mstat model. Both the \Mfid model and the standard \Mstat model (dashed red curve) adopt the baseline radiation efficiency parameters: $f_L=1$, $\beta_L=0$, $f_X=10$, $\beta_X=0$, and spectral index $\alpha=1.5$. The \Mtuned model (dashed blue curve) illustrates a scenario where the efficiencies within the \Mstat framework are explicitly modified ($f_L=10$, $\beta_L=0$, $f_X=5 \times 10^{-4}$, $\beta_X=-3$) to closely reproduce the global absorption trough of the \Mfid model. All theoretical predictions are computed utilizing the \Msink for radiative transfer, with a grid resolution corresponding to $N_\mathrm{grid} = 16$ and a redshift step size of $\Delta z = 0.1$. 
    }
    \label{fig:model_3}
\end{figure}

We next compute the 21-cm signal for the \Mfid and \Mstat models, adopting typical values for the Lyman-$\alpha$ and X-ray efficiencies of $f_L = 1$, $f_X = 10$, respectively, with $\beta_L, \beta_X = 0$, as done in the previous section. The resulting global 21-cm differential brightness temperature, $\delta T_b$, together with its power spectrum, $\Delta^2_{\delta T_b}(k)$, evaluated at $k = 0.05 \,h \,\mathrm{cMpc}^{-1}$, is shown in \fig{fig:model_3}.

As shown in the left panel of \fig{fig:model_3}, the \Mfid model (black curve) exhibits a pronounced absorption trough, reaching a minimum of $\delta T_b \approx -65 \,\mathrm{mK}$ at $z \approx 15$, followed by a smooth transition to an emission peak at $z \approx 9$. In contrast, the \Mstat model (red curve) produces a substantially delayed absorption trough centered at $z \approx 12$. This delay arises from the significant deficit of early galaxies at $z \gtrsim 9$ relative to the \Mfid model, which postpones the onset of both Lyman-$\alpha$ coupling and X-ray heating. At later times, however, the two models exhibit nearly identical emission histories because the ionizing escape efficiency, $\varepsilon_{\rm esc}$, in the \Mstat model has been calibrated to reproduce the fiducial reionization history at $z \sim 6$.

The right panel of \fig{fig:model_3} highlights these differences more clearly through the evolution of the 21-cm power spectrum. The \Mfid model exhibits the familiar three-peak structure, corresponding to fluctuations driven by Lyman-$\alpha$ coupling ($z \approx 18$), X-ray heating ($z \approx 13$), and reionization ($z \approx 7$). In the \Mstat model, the first two peaks are shifted to lower redshifts, with the Lyman-$\alpha$ and X-ray heating peaks occurring at $z \approx 15$ and $z \approx 12$, respectively, reflecting the delayed formation of the first galaxies. In contrast, the reionization peak remains at $z \approx 7$, again owing to the calibration of $\varepsilon_{\rm esc}$. These results demonstrate that both the global 21-cm signal and its spatial fluctuations are highly sensitive probes of the redshift evolution of galaxy properties.

A natural question is whether this signature can be reproduced within the \Mstat model by adjusting the Lyman-$\alpha$ and X-ray efficiencies. We find that varying only the normalization parameters, $f_L$ and $f_X$, is insufficient to reproduce the fiducial 21-cm signal. However, allowing these efficiencies to depend on halo mass, i.e., permitting $\beta_{L}$ and $\beta_{X}$ to differ from zero, yields parameter combinations that closely match the fiducial global signal. Through a simple trial-and-error search, we identify one such model with $f_X = 5 \times 10^{-4}$ and $\beta_{X} = -3$, while keeping the Lyman-$\alpha$ efficiency halo-mass independent with $f_L = 10$ and $\beta_{L} = 0$. The steep X-ray scaling implies that halos with $M_h \lesssim 3.7 \times 10^8 \, M_\odot$ have $f_X \gtrsim 10$, comparable to the fiducial value, whereas more massive halos are significantly less efficient X-ray sources. We call this model \Mtuned.

The resulting global signal for the \Mtuned model is shown in the left panel of \fig{fig:model_3} (blue curve). Increasing the Lyman-$\alpha$ efficiency shifts the absorption feature to earlier times, while the enhanced X-ray efficiency in low-mass halos prevents the absorption trough from becoming excessively deep. Because the X-ray efficiency decreases steeply with halo mass, heating is naturally stronger at high redshifts, when low-mass halos dominate the galaxy population, and weaker at later times, when progressively more massive halos contribute to the emissivity. As a result, the global signal of this \Mtuned model is nearly indistinguishable from that of the \Mfid model.

Despite the remarkable agreement in the global signal, the corresponding 21-cm power spectra differ substantially, as shown in the right panel of \fig{fig:model_3} (blue curve). Although the three characteristic peaks occur at approximately the same redshifts, their amplitudes are systematically lower in the \Mtuned model. This difference reflects the distinct halo populations responsible for the radiation backgrounds. Around the Lyman-$\alpha$ peak ($z \sim 18$), the \Mfid model has an effective star-formation efficiency that scales as $\varepsilon_\star \propto M_h^{2.19}$, whereas the \Mtuned model follows the much shallower relation $\varepsilon_\star \propto M_h^{0.61}$. Consequently, relatively low-mass halos contribute more strongly in the \Mtuned model, reducing the large-scale clustering signal and hence the power-spectrum amplitude.

A similar argument applies during the X-ray heating epoch ($z \sim 13$). In the \Mfid model, $\varepsilon_\star \propto M_h^{1.18}$ while $f_X$ is independent of halo mass. In contrast, the \Mtuned model has $\varepsilon_\star \propto M_h^{0.61}$ together with $f_X \propto M_h^{-3}$, yielding an effective scaling of $M_h^{-2.39}$. X-ray heating is therefore dominated by low-mass halos, which are significantly less clustered than their massive counterparts. As a result, the amplitude of the 21-cm power spectrum remains substantially lower than in the \Mfid model, even though the global signals are nearly identical.

The above comparison demonstrates that the global 21-cm signal alone is insufficient to uniquely constrain the evolution of the underlying galaxy population. Although suitable choices of the Lyman-$\alpha$ and X-ray efficiencies can reproduce the fiducial global signal even within a redshift-independent galaxy model, the corresponding 21-cm power spectrum remains markedly different. This difference arises because the spatial fluctuations retain information about the halo populations responsible for the radiation backgrounds, thereby breaking degeneracies that are unavoidable in global-signal measurements. These results highlight the importance of combining global and fluctuation measurements to robustly constrain the redshift evolution of high-redshift galaxies.

\section{Summary and Conclusion}\label{sec:summary}

In this work, we have presented a major extension of the explicitly photon-conserving semi-numerical framework, \textsf{SCRIPT}, enabling the self-consistent calculation of spin temperature ($T_S$) fluctuations during the Cosmic Dawn. By incorporating inhomogeneous Lyman-$\alpha$ coupling and X-ray heating, the upgraded framework bridges the critical gap between the emergence of the first luminous sources and the later stages of reionization. To ensure numerical rigor, the radiative transfer of Lyman-$\alpha$ and X-ray photons is evaluated using two independent, complementary algorithms (the Source and Sink methods), both of which yield highly consistent results that are demonstrably robust against variations in both spatial and redshift resolution.

We applied this framework to investigate whether the 21-cm signal can independently probe the high-redshift galaxy populations recently uncovered by JWST. Specifically, we compared a \Mfid galaxy formation model, characterized by a redshift-evolving star-formation efficiency tailored to match JWST UVLFs at $z \ge 6$, against a baseline \Mstat model featuring a redshift-independent efficiency.

Our analysis reveals a critical subtlety in the interpretation of Cosmic Dawn observables: the global 21-cm signal is highly degenerate with respect to the underlying astrophysical source properties. We demonstrated that even if the galaxy population remains static, the global absorption trough of a redshift-evolving model can be nearly perfectly replicated by introducing a steep, mass-dependent X-ray efficiency (e.g., $\beta_X = -3$). This tuning artificially shifts the bulk of the heating to highly abundant, low-mass halos, masking the absence of a truly evolving galaxy population.

However, we show that this astrophysical degeneracy is definitively broken by the 21-cm power spectrum. Because the intrinsic spatial fluctuations of the radiation backgrounds are inextricably linked to the clustering bias of their host halos, the power spectrum retains a distinct structural memory of the source population. In our \Mstat scenario, the over-reliance on weakly clustered low-mass halos yields a power spectrum with unambiguously lower amplitudes across all characteristic peaks.

Ultimately, our findings highlight the strong complementarity among different observational probes of the high-redshift Universe (for a recent review, see \citep{2026arXiv260630947C}), an approach that has also been successfully leveraged in a number of previous studies \citep{Park++2019,Qin2020,Munoz2022,Nikolic2024,Bevins2024,Pochinda2024,Dhandha2025,Dhandha2025a,Davies2025}. Global 21-cm measurements provide a vital, physically insightful anchor for the overall cosmic photon budget and the mean thermal history of the IGM. When integrated with the spatial fluctuations probed by low-frequency radio interferometers, this combined approach becomes incredibly powerful. By leveraging the absolute baseline offered by global experiments alongside the structural memory retained in the 21-cm power spectrum, we can robustly break emission degeneracies and directly constrain the faint, early galaxy populations that remain otherwise invisible to direct near-infrared surveys.

Looking forward, the computational efficiency and explicit photon conservation of the extended \textsf{SCRIPT} framework make it an ideal engine for exhaustive parameter space exploration. Future work will focus on leveraging these large-volume simulations to construct artificial neural network emulators. By coupling these fast emulators with advanced machine learning techniques, such as Simulation-Based Inference, we aim to establish a rigorous parameter estimation pipeline. This will be essential for interpreting concurrent data from ongoing global signal experiments and upcoming interferometric facilities like the Square Kilometre Array (SKA), ultimately unlocking the full astrophysical potential of the Cosmic Dawn.

\acknowledgments
The authors acknowledge discussions with Anirban Chakraborty, Saptarshi Sarkar and Pulak Mohapatra.

\section*{Data Availability}

The data generated during this work will be made available upon reasonable request to the corresponding author.

\appendix

\section{Source vs Sink method} \label{app:source_sink}

\begin{figure}[tbp]
    \centering
    \includegraphics[width=1\linewidth]{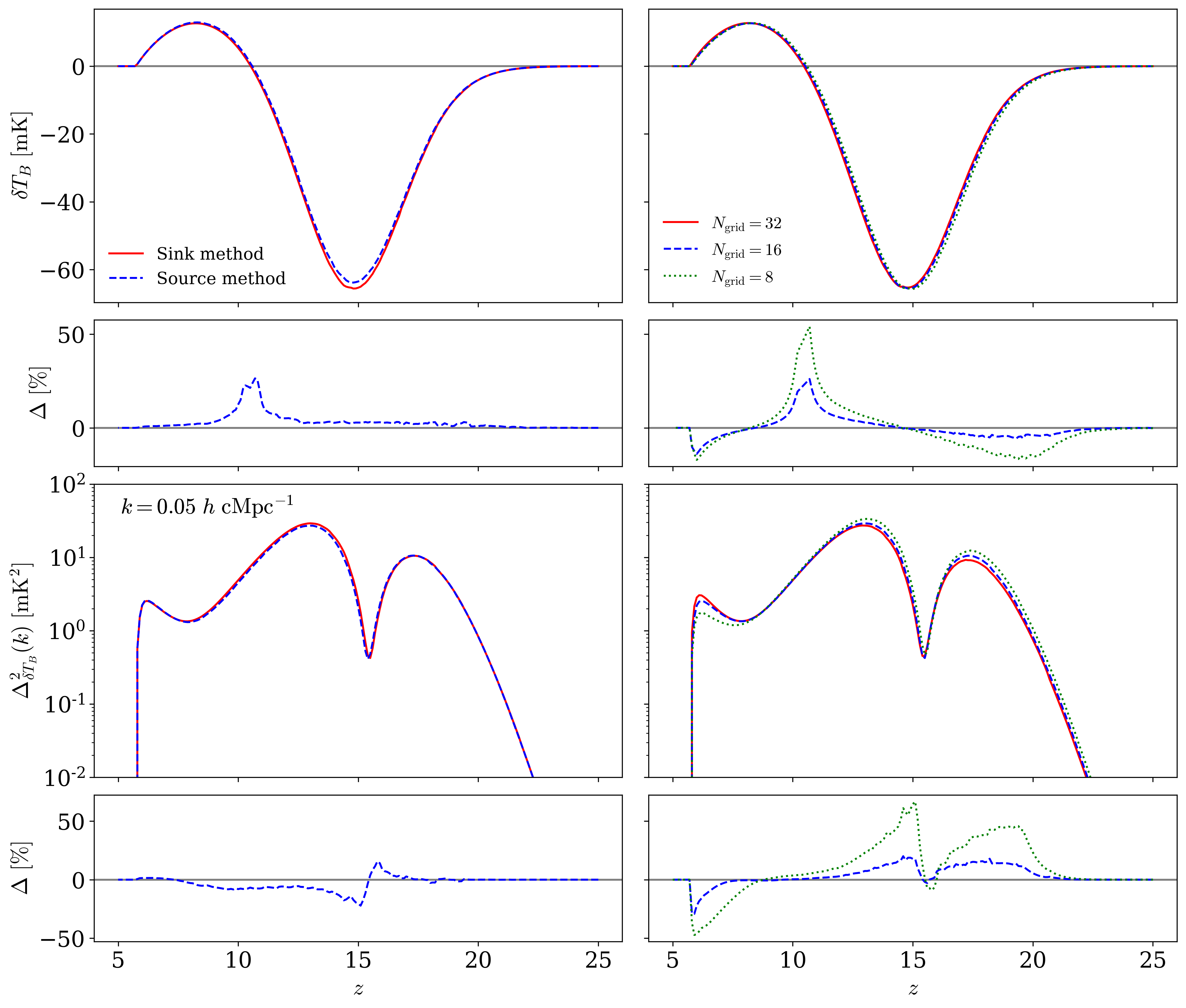}
    \caption{We compare the robustness of our results for different methods and resolutions for global 21-cm signal (top two panels) and its power spectrum (bottom two panels). In the left column we compare the source and sink methods and see that the difference between results is very small. Similarly, in the right column we compare different grid resolutions with $N_\mathrm{grid} = 32, 16, \text{ and } 8$.}
    \label{fig:compare_res_method}
\end{figure}

We have implemented two complementary methods to compute the radiative transfer of Lyman-$\alpha$ and X-ray photons, which we refer to as the \Msink and \Msrc.

In the \Msink, we iterate over sink redshifts $z$ from high to low. At each $z$, for each sink cell, we identify every source shell at higher redshift $z_\mathrm{sr}$ that can contribute photons to this sink cell, and sum over all the contributions. In the \Msrc, we instead iterate over source redshifts $z_\mathrm{sr}$ from high to low. At each $z_\mathrm{sr}$, for each source cell, the photons emitted are propagated outward, into sink cells at lower redshifts $z$. Incorporating the redshift and attenuation processes, we then accumulate the radiation fields and sum over their effects in sink boxes. In both methods, for each sink redshift $z$, photons emitted at higher source redshifts $z_{\rm sr}$ are propagated over intermediate comoving distance bins and these photons are attenuated by distance and intermediate IGM. Their interaction with the medium in the sink cell depends on the temperature, ionization fraction and density of the sink cell.

In \Msink, the field at each $z$ is computed independently and these boxes are saved to disk, after which all source boxes are discarded from RAM. This approach is highly memory-efficient as only the field boxes for a single sink redshift and a single source redshift need to reside in memory simultaneously; however, for each sink redshift, the source boxes must be re-read from disk. In \Msrc, all sink field boxes are held in memory simultaneously throughout the calculation, as we need to accumulate the contributions from all sources before saving the sink boxes. This avoids repeated disk I/O, but requires memory proportional to $N_z \times N_\mathrm{grid}^3$, $N_z$ being the number of redshift snapshots. For the resolutions considered in this work, both methods require approximately similar computation time.

In the continuum limit, the two methods are equivalent. However, in a discretized simulation, there are small differences due to the distance and redshift binning, and the line of sight evolution of intensity due to redshift and intermediate IGM. Results produced from both methods agree reasonably well (Figure~\ref{fig:compare_res_method}, left panels). We use these methods as an independent cross-check. Since they are implemented separately and follow different algorithms, agreement between their output fields validates the implementation of both.

\section{Spatial and Redshift Resolution} \label{app:res}

A primary design requirement of this major update to \textsf{SCRIPT} is that the physical outputs, namely the neutral fraction history, the background radiation intensities, and the resulting 21-cm observables, must remain highly convergent and robust against changes in numerical resolution.

\subsection{Spatial Grid Convergence ($N_\mathrm{grid}$)}

To confirm that our results are independent of box resolution, we evaluate the framework's stability by varying the grid cell counts ($N_\mathrm{grid} = 32, 16, \text{and } 8$) inside a fixed comoving volume. The right panel of \fig{fig:compare_res_method} presents the mean 21-cm signal and the large-scale power spectrum at $k \simeq 0.05~h\,\uni{cMpc}{-1}$ for these spatial configurations, along with their respective percentage differences.

Even at a coarse resolution of $N_\mathrm{grid} = 8$, the overall shape of the signal matches the finer grids remarkably well, with noticeable percentage differences occurring only where the signal approaches zero ($\delta T_b \approx 0$). Meanwhile, the intermediate $N_\mathrm{grid} = 16$ provides an almost exact match to the $N_\mathrm{grid} = 32$ baseline throughout the entire timeline.

Crucially, this large-scale consistency is maintained primarily due to the introduction of the cell-horizon weighting factor $w_{\rm shell}(m)$ detailed in \eqn{eq:w_shell}. Because individual Lyman-$m$ line horizons can be significantly smaller than the comoving width of a coarse cell ($\Delta r_c$), failing to scale the local volume contributions would lead to an unphysical over-counting of injected photons relative to continuum background photons. The implementation of $w_{\rm shell}(m)$ successfully standardizes photon deposition across all grid choices, rendering the code robust and perfectly optimized for high-throughput statistical modeling frameworks.

\subsection{Redshift Integrator Convergence ($\Delta z$)}

While the main results presented in this work utilize a fine redshift step of $\Delta z = 0.1$, we also evaluate the temporal stability of our radiative transfer integration schemes by testing a coarser, computationally optimized resolution of $\Delta z = 1$. We find that our outputs remain highly robust even at this coarse temporal resolution (not shown here). This exceptional temporal alignment indicates that our discrete time-stepping algorithm handles the evolution of Lyman-$\alpha$ coupling and X-ray heating fields without introducing numerical diffusion or integration drift, validating the potential use of larger redshift steps for rapid parameter space exploration.

\providecommand{\href}[2]{#2}\begingroup\raggedright\endgroup

\begin{thebibliography}{100}

\bibitem{Naidu2022}
R.P.~{Naidu}, P.A.~{Oesch}, P.~{van Dokkum}, E.J.~{Nelson}, K.A.~{Suess},
  G.~{Brammer} et~al., \emph{{Two Remarkably Luminous Galaxy Candidates at z
  {\ensuremath{\approx}} 10-12 Revealed by JWST}},
  \href{https://doi.org/10.3847/2041-8213/ac9b22}{\emph{\apjl} {\bfseries 940}
  (2022) L14} [\href{https://arxiv.org/abs/2207.09434}{{\ttfamily
  2207.09434}}].

\bibitem{Castellano2022}
M.~{Castellano}, A.~{Fontana}, T.~{Treu}, P.~{Santini}, E.~{Merlin},
  N.~{Leethochawalit} et~al., \emph{{Early Results from GLASS-JWST. III. Galaxy
  Candidates at z 9-15}},
  \href{https://doi.org/10.3847/2041-8213/ac94d0}{\emph{\apjl} {\bfseries 938}
  (2022) L15} [\href{https://arxiv.org/abs/2207.09436}{{\ttfamily
  2207.09436}}].

\bibitem{Finkelstein2022}
S.L.~{Finkelstein}, M.B.~{Bagley}, P.~{Arrabal Haro}, M.~{Dickinson},
  H.C.~{Ferguson}, J.S.~{Kartaltepe} et~al., \emph{{A Long Time Ago in a Galaxy
  Far, Far Away: A Candidate z {\ensuremath{\sim}} 12 Galaxy in Early JWST
  CEERS Imaging}}, \href{https://doi.org/10.3847/2041-8213/ac966e}{\emph{\apjl}
  {\bfseries 940} (2022) L55}
  [\href{https://arxiv.org/abs/2207.12474}{{\ttfamily 2207.12474}}].

\bibitem{Labbe2023}
I.~{Labb{\'e}}, P.~{van Dokkum}, E.~{Nelson}, R.~{Bezanson}, K.A.~{Suess},
  J.~{Leja} et~al., \emph{{A population of red candidate massive galaxies 600
  Myr after the Big Bang}},
  \href{https://doi.org/10.1038/s41586-023-05786-2}{\emph{\nat} {\bfseries 616}
  (2023) 266} [\href{https://arxiv.org/abs/2207.12446}{{\ttfamily
  2207.12446}}].

\bibitem{Atek2023}
H.~{Atek}, M.~{Shuntov}, L.J.~{Furtak}, J.~{Richard}, J.-P.~{Kneib},
  G.~{Mahler} et~al., \emph{{Revealing galaxy candidates out to z 16 with JWST
  observations of the lensing cluster SMACS0723}},
  \href{https://doi.org/10.1093/mnras/stac3144}{\emph{\mnras} {\bfseries 519}
  (2023) 1201} [\href{https://arxiv.org/abs/2207.12338}{{\ttfamily
  2207.12338}}].

\bibitem{Adams2023}
N.J.~{Adams}, C.J.~{Conselice}, L.~{Ferreira}, D.~{Austin}, J.A.A.~{Trussler},
  I.~{Juod{\v{z}}balis} et~al., \emph{{Discovery and properties of ultra-high
  redshift galaxies (9 < z < 12) in the JWST ERO SMACS 0723 Field}},
  \href{https://doi.org/10.1093/mnras/stac3347}{\emph{\mnras} {\bfseries 518}
  (2023) 4755} [\href{https://arxiv.org/abs/2207.11217}{{\ttfamily
  2207.11217}}].

\bibitem{Bradley2023}
L.D.~{Bradley}, D.~{Coe}, G.~{Brammer}, L.J.~{Furtak}, R.L.~{Larson},
  V.~{Kokorev} et~al., \emph{{High-redshift Galaxy Candidates at z = 9-10 as
  Revealed by JWST Observations of WHL0137-08}},
  \href{https://doi.org/10.3847/1538-4357/acecfe}{\emph{\apj} {\bfseries 955}
  (2023) 13} [\href{https://arxiv.org/abs/2210.01777}{{\ttfamily 2210.01777}}].

\bibitem{Whitler2023}
L.~{Whitler}, R.~{Endsley}, D.P.~{Stark}, M.~{Topping}, Z.~{Chen} and
  S.~{Charlot}, \emph{{On the ages of bright galaxies 500 Myr after the big
  bang: insights into star formation activity at z {\ensuremath{\gtrsim}} 15
  with JWST}}, \href{https://doi.org/10.1093/mnras/stac3535}{\emph{\mnras}
  {\bfseries 519} (2023) 157}
  [\href{https://arxiv.org/abs/2208.01599}{{\ttfamily 2208.01599}}].

\bibitem{Robertson2024}
B.~{Robertson}, B.D.~{Johnson}, S.~{Tacchella}, D.J.~{Eisenstein},
  K.~{Hainline}, S.~{Arribas} et~al., \emph{{Earliest Galaxies in the JADES
  Origins Field: Luminosity Function and Cosmic Star Formation Rate Density 300
  Myr after the Big Bang}},
  \href{https://doi.org/10.3847/1538-4357/ad463d}{\emph{\apj} {\bfseries 970}
  (2024) 31} [\href{https://arxiv.org/abs/2312.10033}{{\ttfamily 2312.10033}}].

\bibitem{Castellano2023_UVLF}
M.~{Castellano}, A.~{Fontana}, T.~{Treu}, E.~{Merlin}, P.~{Santini},
  P.~{Bergamini} et~al., \emph{{Early Results from GLASS-JWST. XIX. A High
  Density of Bright Galaxies at z {\ensuremath{\approx}} 10 in the A2744
  Region}}, \href{https://doi.org/10.3847/2041-8213/accea5}{\emph{\apjl}
  {\bfseries 948} (2023) L14}
  [\href{https://arxiv.org/abs/2212.06666}{{\ttfamily 2212.06666}}].

\bibitem{Finkelstein2023_UVLF}
S.L.~{Finkelstein}, M.B.~{Bagley}, H.C.~{Ferguson}, S.M.~{Wilkins},
  J.S.~{Kartaltepe}, C.~{Papovich} et~al., \emph{{CEERS Key Paper. I. An Early
  Look into the First 500 Myr of Galaxy Formation with JWST}},
  \href{https://doi.org/10.3847/2041-8213/acade4}{\emph{\apjl} {\bfseries 946}
  (2023) L13} [\href{https://arxiv.org/abs/2211.05792}{{\ttfamily
  2211.05792}}].

\bibitem{Gonzalez2023}
P.G.~{P{\'e}rez-Gonz{\'a}lez}, L.~{Costantin}, D.~{Langeroodi}, P.~{Rinaldi},
  M.~{Annunziatella}, O.~{Ilbert} et~al., \emph{{Life beyond 30: Probing the
  -20 < M $_{UV}$ < -17 Luminosity Function at 8 < z < 13 with the NIRCam
  Parallel Field of the MIRI Deep Survey}},
  \href{https://doi.org/10.3847/2041-8213/acd9d0}{\emph{\apjl} {\bfseries 951}
  (2023) L1} [\href{https://arxiv.org/abs/2302.02429}{{\ttfamily 2302.02429}}].

\bibitem{Donnan2023}
C.T.~{Donnan}, D.J.~{McLeod}, J.S.~{Dunlop}, R.J.~{McLure}, A.C.~{Carnall},
  R.~{Begley} et~al., \emph{{The evolution of the galaxy UV luminosity function
  at redshifts z = 8 - 15 from deep JWST and ground-based near-infrared
  imaging}}, \href{https://doi.org/10.1093/mnras/stac3472}{\emph{\mnras}
  {\bfseries 518} (2023) 6011}
  [\href{https://arxiv.org/abs/2207.12356}{{\ttfamily 2207.12356}}].

\bibitem{Harikane2023}
Y.~{Harikane}, M.~{Ouchi}, M.~{Oguri}, Y.~{Ono}, K.~{Nakajima}, Y.~{Isobe}
  et~al., \emph{{A Comprehensive Study of Galaxies at z 9-16 Found in the Early
  JWST Data: Ultraviolet Luminosity Functions and Cosmic Star Formation History
  at the Pre-reionization Epoch}},
  \href{https://doi.org/10.3847/1538-4365/acaaa9}{\emph{\apjs} {\bfseries 265}
  (2023) 5} [\href{https://arxiv.org/abs/2208.01612}{{\ttfamily 2208.01612}}].

\bibitem{Bouwens2023}
R.~{Bouwens}, G.~{Illingworth}, P.~{Oesch}, M.~{Stefanon}, R.~{Naidu}, I.~{van
  Leeuwen} et~al., \emph{{UV luminosity density results at z > 8 from the first
  JWST/NIRCam fields: limitations of early data sets and the need for
  spectroscopy}}, \href{https://doi.org/10.1093/mnras/stad1014}{\emph{\mnras}
  {\bfseries 523} (2023) 1009}
  [\href{https://arxiv.org/abs/2212.06683}{{\ttfamily 2212.06683}}].

\bibitem{McLeod2024}
D.J.~{McLeod}, C.T.~{Donnan}, R.J.~{McLure}, J.S.~{Dunlop}, D.~{Magee},
  R.~{Begley} et~al., \emph{{The galaxy UV luminosity function at z $\simeq$ 11
  from a suite of public JWST ERS, ERO, and Cycle-1 programs}},
  \href{https://doi.org/10.1093/mnras/stad3471}{\emph{\mnras} {\bfseries 527}
  (2024) 5004} [\href{https://arxiv.org/abs/2304.14469}{{\ttfamily
  2304.14469}}].

\bibitem{Adams2024}
N.J.~{Adams}, C.J.~{Conselice}, D.~{Austin}, T.~{Harvey}, L.~{Ferreira},
  J.~{Trussler} et~al., \emph{{EPOCHS. II. The Ultraviolet Luminosity Function
  from 7.5 < z < 13.5 Using 180 arcmin$^{2}$ of Deep, Blank Fields from the
  PEARLS Survey and Public JWST Data}},
  \href{https://doi.org/10.3847/1538-4357/ad2a7b}{\emph{\apj} {\bfseries 965}
  (2024) 169} [\href{https://arxiv.org/abs/2304.13721}{{\ttfamily
  2304.13721}}].

\bibitem{Finkelstein2024_UVLF}
S.L.~{Finkelstein}, G.C.K.~{Leung}, M.B.~{Bagley}, M.~{Dickinson},
  H.C.~{Ferguson}, C.~{Papovich} et~al., \emph{{The Complete CEERS Early
  Universe Galaxy Sample: A Surprisingly Slow Evolution of the Space Density of
  Bright Galaxies at z {\ensuremath{\sim}} 8.5{\textendash}14.5}},
  \href{https://doi.org/10.3847/2041-8213/ad4495}{\emph{\apjl} {\bfseries 969}
  (2024) L2} [\href{https://arxiv.org/abs/2311.04279}{{\ttfamily 2311.04279}}].

\bibitem{Whitler2025}
L.~{Whitler}, D.P.~{Stark}, M.W.~{Topping}, B.~{Robertson}, M.~{Rieke},
  K.N.~{Hainline} et~al., \emph{{The $z rsim 9$ galaxy UV luminosity function
  from the JWST Advanced Deep Extragalactic Survey: insights into early galaxy
  evolution and reionization}},
  \href{https://doi.org/10.48550/arXiv.2501.00984}{\emph{arXiv e-prints} (2025)
  arXiv:2501.00984} [\href{https://arxiv.org/abs/2501.00984}{{\ttfamily
  2501.00984}}].

\bibitem{Gonzalez2025}
P.G.~{P{\'e}rez-Gonz{\'a}lez}, G.~{{\"O}stlin}, L.~{Costantin}, J.~{Melinder},
  S.L.~{Finkelstein}, R.S.~{Somerville} et~al., \emph{{The rise of the galactic
  empire: luminosity functions at $z\sim17$ and $z\sim25$ estimated with the
  MIDIS$+$NGDEEP ultra-deep JWST/NIRCam dataset}},
  \href{https://doi.org/10.48550/arXiv.2503.15594}{\emph{arXiv e-prints} (2025)
  arXiv:2503.15594} [\href{https://arxiv.org/abs/2503.15594}{{\ttfamily
  2503.15594}}].

\bibitem{Weibel2025}
A.~{Weibel}, P.A.~{Oesch}, C.C.~{Williams}, C.K.~{Jespersen}, M.~{Shuntov},
  K.E.~{Whitaker} et~al., \emph{{Exploring Cosmic Dawn with PANORAMIC I: The
  Bright End of the UVLF at $z\sim9 -17$}},
  \href{https://doi.org/10.48550/arXiv.2507.06292}{\emph{arXiv e-prints} (2025)
  arXiv:2507.06292} [\href{https://arxiv.org/abs/2507.06292}{{\ttfamily
  2507.06292}}].

\bibitem{Dekel2023}
A.~{Dekel}, K.S.~{Sarkar}, Y.~{Birnboim}, N.~{Mandelker} and Z.~{Li},
  \emph{{Efficient Formation of Massive Galaxies at Cosmic Dawn by
  Feedback-Free Starbursts}},
  \href{https://doi.org/10.48550/arXiv.2303.04827}{\emph{arXiv e-prints} (2023)
  arXiv:2303.04827} [\href{https://arxiv.org/abs/2303.04827}{{\ttfamily
  2303.04827}}].

\bibitem{Li2024}
Z.~{Li}, A.~{Dekel}, K.C.~{Sarkar}, H.~{Aung}, M.~{Giavalisco}, N.~{Mandelker}
  et~al., \emph{{Feedback-free starbursts at cosmic dawn: Observable
  predictions for JWST}},
  \href{https://doi.org/10.1051/0004-6361/202348727}{\emph{\aap} {\bfseries
  690} (2024) A108} [\href{https://arxiv.org/abs/2311.14662}{{\ttfamily
  2311.14662}}].

\bibitem{Qin2023}
Y.~{Qin}, S.~{Balu} and J.S.B.~{Wyithe}, \emph{{Implications of z
  {\ensuremath{\gtrsim}} 12 JWST galaxies for galaxy formation at high
  redshift}}, \href{https://doi.org/10.1093/mnras/stad2448}{\emph{\mnras}
  {\bfseries 526} (2023) 1324}
  [\href{https://arxiv.org/abs/2305.17959}{{\ttfamily 2305.17959}}].

\bibitem{Chakraborty2024}
A.~{Chakraborty} and T.R.~{Choudhury}, \emph{{Modelling the star-formation
  activity and ionizing properties of high-redshift galaxies}},
  \href{https://doi.org/10.1088/1475-7516/2024/07/078}{\emph{\jcap} {\bfseries
  2024} (2024) 078} [\href{https://arxiv.org/abs/2404.02879}{{\ttfamily
  2404.02879}}].

\bibitem{Chakraborty2026}
A.~{Chakraborty} and T.R.~{Choudhury}, \emph{{Probing reionization-era galaxies
  with JWST UV luminosity functions and large-scale clustering}},
  \href{https://doi.org/10.1088/1475-7516/2026/01/008}{\emph{\jcap} {\bfseries
  2026} (2026) 008} [\href{https://arxiv.org/abs/2503.07590}{{\ttfamily
  2503.07590}}].

\bibitem{Mirocha2023}
J.~{Mirocha} and S.R.~{Furlanetto}, \emph{{Balancing the efficiency and
  stochasticity of star formation with dust extinction in $z > 10$ galaxies
  observed by JWST}},
  \href{https://doi.org/10.1093/mnras/stac3578}{\emph{\mnras} {\bfseries 519}
  (2023) 843} [\href{https://arxiv.org/abs/2208.12826}{{\ttfamily
  2208.12826}}].

\bibitem{Mason2023}
C.A.~{Mason}, M.~{Trenti} and T.~{Treu}, \emph{{The brightest galaxies at
  cosmic dawn}}, \href{https://doi.org/10.1093/mnras/stad035}{\emph{\mnras}
  {\bfseries 521} (2023) 497}
  [\href{https://arxiv.org/abs/2207.14808}{{\ttfamily 2207.14808}}].

\bibitem{Shen2023}
X.~{Shen}, M.~{Vogelsberger}, M.~{Boylan-Kolchin}, S.~{Tacchella} and
  R.~{Kannan}, \emph{{The impact of UV variability on the abundance of bright
  galaxies at z {\ensuremath{\geq}} 9}},
  \href{https://doi.org/10.1093/mnras/stad2508}{\emph{\mnras} {\bfseries 525}
  (2023) 3254} [\href{https://arxiv.org/abs/2305.05679}{{\ttfamily
  2305.05679}}].

\bibitem{Sun2023}
G.~{Sun}, C.-A.~{Faucher-Gigu{\`e}re}, C.C.~{Hayward}, X.~{Shen}, A.~{Wetzel}
  and R.K.~{Cochrane}, \emph{{Bursty Star Formation Naturally Explains the
  Abundance of Bright Galaxies at Cosmic Dawn}},
  \href{https://doi.org/10.3847/2041-8213/acf85a}{\emph{\apjl} {\bfseries 955}
  (2023) L35} [\href{https://arxiv.org/abs/2307.15305}{{\ttfamily
  2307.15305}}].

\bibitem{Pallottini2023}
A.~{Pallottini} and A.~{Ferrara}, \emph{{Stochastic star formation in early
  galaxies: JWST implications}},
  \href{https://doi.org/10.48550/arXiv.2307.03219}{\emph{arXiv e-prints} (2023)
  arXiv:2307.03219} [\href{https://arxiv.org/abs/2307.03219}{{\ttfamily
  2307.03219}}].

\bibitem{Gelli2024}
V.~{Gelli}, C.~{Mason} and C.C.~{Hayward}, \emph{{The Impact of Mass-dependent
  Stochasticity at Cosmic Dawn}},
  \href{https://doi.org/10.3847/1538-4357/ad7b36}{\emph{\apj} {\bfseries 975}
  (2024) 192} [\href{https://arxiv.org/abs/2405.13108}{{\ttfamily
  2405.13108}}].

\bibitem{Kravtsov2024}
A.~{Kravtsov} and V.~{Belokurov}, \emph{{Stochastic star formation and the
  abundance of $z>10$ UV-bright galaxies}},
  \href{https://doi.org/10.48550/arXiv.2405.04578}{\emph{arXiv e-prints} (2024)
  arXiv:2405.04578} [\href{https://arxiv.org/abs/2405.04578}{{\ttfamily
  2405.04578}}].

\bibitem{Inayoshi2022}
K.~{Inayoshi}, Y.~{Harikane}, A.K.~{Inoue}, W.~{Li} and L.C.~{Ho}, \emph{{A
  Lower Bound of Star Formation Activity in Ultra-high-redshift Galaxies
  Detected with JWST: Implications for Stellar Populations and Radiation
  Sources}}, \href{https://doi.org/10.3847/2041-8213/ac9310}{\emph{\apjl}
  {\bfseries 938} (2022) L10}
  [\href{https://arxiv.org/abs/2208.06872}{{\ttfamily 2208.06872}}].

\bibitem{Trinca2023}
A.~{Trinca}, R.~{Schneider}, R.~{Valiante}, L.~{Graziani}, A.~{Ferrotti},
  K.~{Omukai} et~al., \emph{{Exploring the nature of UV-bright $z rsim 10$
  galaxies detected by JWST: star formation, black hole accretion, or a
  non-universal IMF?}},
  \href{https://doi.org/10.48550/arXiv.2305.04944}{\emph{arXiv e-prints} (2023)
  arXiv:2305.04944} [\href{https://arxiv.org/abs/2305.04944}{{\ttfamily
  2305.04944}}].

\bibitem{Ventura2024}
E.M.~{Ventura}, Y.~{Qin}, S.~{Balu} and J.S.B.~{Wyithe}, \emph{{Semi-analytic
  modelling of Pop. III star formation and metallicity evolution - I. Impact on
  the UV luminosity functions at z = 9-16}},
  \href{https://doi.org/10.1093/mnras/stae567}{\emph{\mnras} {\bfseries 529}
  (2024) 628} [\href{https://arxiv.org/abs/2401.07396}{{\ttfamily
  2401.07396}}].

\bibitem{Yung2024}
L.Y.A.~{Yung}, R.S.~{Somerville}, S.L.~{Finkelstein}, S.M.~{Wilkins} and
  J.P.~{Gardner}, \emph{{Are the ultra-high-redshift galaxies at z > 10
  surprising in the context of standard galaxy formation models?}},
  \href{https://doi.org/10.1093/mnras/stad3484}{\emph{\mnras} {\bfseries 527}
  (2024) 5929} [\href{https://arxiv.org/abs/2304.04348}{{\ttfamily
  2304.04348}}].

\bibitem{Hutter2025}
A.~{Hutter}, E.R.~{Cueto}, P.~{Dayal}, S.~{Gottl{\"o}ber}, M.~{Trebitsch} and
  G.~{Yepes}, \emph{{ASTRAEUS: X. Indications of a top-heavy initial mass
  function in highly star-forming galaxies from JWST observations at z > 10}},
  \href{https://doi.org/10.1051/0004-6361/202452460}{\emph{\aap} {\bfseries
  694} (2025) A254} [\href{https://arxiv.org/abs/2410.00730}{{\ttfamily
  2410.00730}}].

\bibitem{Lu2025}
S.~{Lu}, C.S.~{Frenk}, S.~{Bose}, C.G.~{Lacey}, S.~{Cole}, C.M.~{Baugh} et~al.,
  \emph{{A comparison of pre-existing {\ensuremath{\Lambda}}CDM predictions
  with the abundance of JWST galaxies at high redshift}},
  \href{https://doi.org/10.1093/mnras/stae2646}{\emph{\mnras} {\bfseries 536}
  (2025) 1018} [\href{https://arxiv.org/abs/2406.02672}{{\ttfamily
  2406.02672}}].

\bibitem{Ferrara2023}
A.~{Ferrara}, A.~{Pallottini} and P.~{Dayal}, \emph{{On the stunning abundance
  of super-early, luminous galaxies revealed by JWST}},
  \href{https://doi.org/10.1093/mnras/stad1095}{\emph{\mnras} {\bfseries 522}
  (2023) 3986} [\href{https://arxiv.org/abs/2208.00720}{{\ttfamily
  2208.00720}}].

\bibitem{Ziparo2023}
F.~{Ziparo}, A.~{Ferrara}, L.~{Sommovigo} and M.~{Kohandel}, \emph{{Blue
  monsters. Why are JWST super-early, massive galaxies so blue?}},
  \href{https://doi.org/10.1093/mnras/stad125}{\emph{\mnras} {\bfseries 520}
  (2023) 2445} [\href{https://arxiv.org/abs/2209.06840}{{\ttfamily
  2209.06840}}].

\bibitem{Ferrara2024}
A.~{Ferrara}, \emph{{Super-early JWST galaxies, outflows, and
  Ly{\ensuremath{\alpha}} visibility in the Epoch of Reionization}},
  \href{https://doi.org/10.1051/0004-6361/202348321}{\emph{\aap} {\bfseries
  684} (2024) A207} [\href{https://arxiv.org/abs/2310.12197}{{\ttfamily
  2310.12197}}].

\bibitem{Ferrara2025}
A.~{Ferrara}, A.~{Pallottini} and L.~{Sommovigo}, \emph{{Blue monsters at z >
  10: Where all their dust has gone}},
  \href{https://doi.org/10.1051/0004-6361/202452707}{\emph{\aap} {\bfseries
  694} (2025) A286} [\href{https://arxiv.org/abs/2410.19042}{{\ttfamily
  2410.19042}}].

\bibitem{Pacucci2022}
F.~{Pacucci}, P.~{Dayal}, Y.~{Harikane}, A.K.~{Inoue} and A.~{Loeb}, \emph{{Are
  the newly-discovered z 13 drop-out sources starburst galaxies or quasars?}},
  \href{https://doi.org/10.1093/mnrasl/slac035}{\emph{\mnras} {\bfseries 514}
  (2022) L6} [\href{https://arxiv.org/abs/2201.00823}{{\ttfamily 2201.00823}}].

\bibitem{Hegde2024}
S.~{Hegde}, M.M.~{Wyatt} and S.R.~{Furlanetto}, \emph{{A hidden population of
  active galactic nuclei can explain the overabundance of luminous z > 10
  objects observed by JWST}},
  \href{https://doi.org/10.1088/1475-7516/2024/08/025}{\emph{\jcap} {\bfseries
  2024} (2024) 025} [\href{https://arxiv.org/abs/2405.01629}{{\ttfamily
  2405.01629}}].

\bibitem{Fujimoto2024}
S.~{Fujimoto}, B.~{Wang}, J.R.~{Weaver}, V.~{Kokorev}, H.~{Atek}, R.~{Bezanson}
  et~al., \emph{{UNCOVER: A NIRSpec Census of Lensed Galaxies at z =
  8.50{\textendash}13.08 Probing a High-AGN Fraction and Ionized Bubbles in the
  Shadow}}, \href{https://doi.org/10.3847/1538-4357/ad9027}{\emph{\apj}
  {\bfseries 977} (2024) 250}
  [\href{https://arxiv.org/abs/2308.11609}{{\ttfamily 2308.11609}}].

\bibitem{Bowman2018}
J.D.~{Bowman}, A.E.E.~{Rogers}, R.A.~{Monsalve}, T.J.~{Mozdzen} and
  N.~{Mahesh}, \emph{{An absorption profile centred at 78 megahertz in the
  sky-averaged spectrum}},
  \href{https://doi.org/10.1038/nature25792}{\emph{\nat} {\bfseries 555} (2018)
  67} [\href{https://arxiv.org/abs/1810.05912}{{\ttfamily 1810.05912}}].

\bibitem{Singh2022}
S.~{Singh}, N.T.~{Jishnu}, R.~{Subrahmanyan}, N.~{Udaya Shankar},
  B.S.~{Girish}, A.~{Raghunathan} et~al., \emph{{On the detection of a cosmic
  dawn signal in the radio background}},
  \href{https://doi.org/10.1038/s41550-022-01610-5}{\emph{Nature Astronomy}
  {\bfseries 6} (2022) 607} [\href{https://arxiv.org/abs/2112.06778}{{\ttfamily
  2112.06778}}].

\bibitem{deLeraAcedo2022}
E.~{de Lera Acedo}, D.I.L.~{de Villiers}, N.~{Razavi-Ghods}, W.~{Handley},
  A.~{Fialkov}, A.~{Magro} et~al., \emph{{The REACH radiometer for detecting
  the 21-cm hydrogen signal from redshift z {\ensuremath{\approx}} 7.5-28}},
  \href{https://doi.org/10.1038/s41550-022-01709-9}{\emph{Nature Astronomy}
  {\bfseries 6} (2022) 984} [\href{https://arxiv.org/abs/2210.07409}{{\ttfamily
  2210.07409}}].

\bibitem{Monsalve2024}
R.A.~{Monsalve}, C.~{Altamirano}, V.~{Bidula}, R.~{Bustos}, C.H.~{Bye},
  H.C.~{Chiang} et~al., \emph{{Mapper of the IGM spin temperature: instrument
  overview}}, \href{https://doi.org/10.1093/mnras/stae1138}{\emph{\mnras}
  {\bfseries 530} (2024) 4125}
  [\href{https://arxiv.org/abs/2309.02996}{{\ttfamily 2309.02996}}].

\bibitem{2025RASTI...4af046B}
P.~{Bull}, A.~{El-Makadema}, H.~{Garsden}, J.~{Edgley}, N.~{Roddis},
  J.~{Chluba} et~al., \emph{{RHINO: a large horn antenna for detecting the 21
  cm global signal}}, \href{https://doi.org/10.1093/rasti/rzaf046}{\emph{RAS
  Techniques and Instruments} {\bfseries 4} (2025) rzaf046}
  [\href{https://arxiv.org/abs/2410.00076}{{\ttfamily 2410.00076}}].

\bibitem{Philip2019}
L.~{Philip}, Z.~{Abdurashidova}, H.C.~{Chiang}, N.~{Ghazi}, A.~{Gumba},
  H.M.~{Heilgendorff} et~al., \emph{{Probing Radio Intensity at High-Z from
  Marion: 2017 Instrument}},
  \href{https://doi.org/10.1142/S2251171719500041}{\emph{Journal of
  Astronomical Instrumentation} {\bfseries 8} (2019) 1950004}
  [\href{https://arxiv.org/abs/1806.09531}{{\ttfamily 1806.09531}}].

\bibitem{Paciga2013}
G.~{Paciga}, J.G.~{Albert}, K.~{Bandura}, T.-C.~{Chang}, Y.~{Gupta},
  C.~{Hirata} et~al., \emph{{A simulation-calibrated limit on the H I power
  spectrum from the GMRT Epoch of Reionization experiment}},
  \href{https://doi.org/10.1093/mnras/stt753}{\emph{\mnras} {\bfseries 433}
  (2013) 639} [\href{https://arxiv.org/abs/1301.5906}{{\ttfamily 1301.5906}}].

\bibitem{Tingay2013}
S.J.~{Tingay}, R.~{Goeke}, J.D.~{Bowman}, D.~{Emrich}, S.M.~{Ord},
  D.A.~{Mitchell} et~al., \emph{{The Murchison Widefield Array: The Square
  Kilometre Array Precursor at Low Radio Frequencies}},
  \href{https://doi.org/10.1017/pasa.2012.007}{\emph{\pasa} {\bfseries 30}
  (2013) e007} [\href{https://arxiv.org/abs/1206.6945}{{\ttfamily 1206.6945}}].

\bibitem{vanHaarlem2013}
M.P.~{van Haarlem}, M.W.~{Wise}, A.W.~{Gunst}, G.~{Heald}, J.P.~{McKean},
  J.W.T.~{Hessels} et~al., \emph{{LOFAR: The LOw-Frequency ARray}},
  \href{https://doi.org/10.1051/0004-6361/201220873}{\emph{\aap} {\bfseries
  556} (2013) A2} [\href{https://arxiv.org/abs/1305.3550}{{\ttfamily
  1305.3550}}].

\bibitem{DeBoer2017}
D.R.~DeBoer, A.R.~Parsons, J.E.~Aguirre, P.~Alexander, Z.S.~Ali, A.P.~Beardsley
  et~al., \emph{Hydrogen epoch of reionization array (hera)},
  \href{https://doi.org/10.1088/1538-3873/129/974/045001}{\emph{Publications of
  the Astronomical Society of the Pacific} {\bfseries 129} (2017) 045001}.

\bibitem{2026arXiv260626435C}
{CD Science Working Group}, G.~{Bernardi}, D.~{Breitman}, A.~{Datta},
  A.~{Fialkov}, L.V.E.~{Koopmans} et~al., \emph{{Overview of 21cm Experiments
  at high redshift with SKAO}},
  \href{https://doi.org/10.48550/arXiv.2606.26435}{\emph{arXiv e-prints} (2026)
  arXiv:2606.26435} [\href{https://arxiv.org/abs/2606.26435}{{\ttfamily
  2606.26435}}].

\bibitem{2006PhR...433..181F}
S.R.~{Furlanetto}, S.P.~{Oh} and F.H.~{Briggs}, \emph{{Cosmology at low
  frequencies: The 21 cm transition and the high-redshift Universe}},
  \href{https://doi.org/10.1016/j.physrep.2006.08.002}{\emph{\physrep}
  {\bfseries 433} (2006) 181}
  [\href{https://arxiv.org/abs/astro-ph/0608032}{{\ttfamily
  astro-ph/0608032}}].

\bibitem{Pritchard2012}
J.R.~{Pritchard} and A.~{Loeb}, \emph{{21 cm cosmology in the 21st century}},
  \href{https://doi.org/10.1088/0034-4885/75/8/086901}{\emph{Reports on
  Progress in Physics} {\bfseries 75} (2012) 086901}
  [\href{https://arxiv.org/abs/1109.6012}{{\ttfamily 1109.6012}}].

\bibitem{Park++2019}
J.~{Park}, A.~{Mesinger}, B.~{Greig} and N.~{Gillet}, \emph{{Inferring the
  astrophysics of reionization and cosmic dawn from galaxy luminosity functions
  and the 21-cm signal}},
  \href{https://doi.org/10.1093/mnras/stz032}{\emph{\mnras} {\bfseries 484}
  (2019) 933} [\href{https://arxiv.org/abs/1809.08995}{{\ttfamily
  1809.08995}}].

\bibitem{Qin2020}
Y.~{Qin}, A.~{Mesinger}, J.~{Park}, B.~{Greig} and J.B.~{Mu{\~n}oz}, \emph{{A
  tale of two sites - I. Inferring the properties of minihalo-hosted galaxies
  from current observations}},
  \href{https://doi.org/10.1093/mnras/staa1131}{\emph{\mnras} {\bfseries 495}
  (2020) 123} [\href{https://arxiv.org/abs/2003.04442}{{\ttfamily
  2003.04442}}].

\bibitem{Munoz2022}
J.B.~{Mu{\~n}oz}, Y.~{Qin}, A.~{Mesinger}, S.G.~{Murray}, B.~{Greig} and
  C.~{Mason}, \emph{{The impact of the first galaxies on cosmic dawn and
  reionization}}, \href{https://doi.org/10.1093/mnras/stac185}{\emph{\mnras}
  {\bfseries 511} (2022) 3657}
  [\href{https://arxiv.org/abs/2110.13919}{{\ttfamily 2110.13919}}].

\bibitem{Nikolic2024}
I.~{Nikoli{\'c}}, A.~{Mesinger}, J.E.~{Davies} and D.~{Prelogovi{\'c}},
  \emph{{The importance of stochasticity in determining galaxy emissivities and
  UV LFs during cosmic dawn and reionization}},
  \href{https://doi.org/10.1051/0004-6361/202451213}{\emph{\aap} {\bfseries
  692} (2024) A142} [\href{https://arxiv.org/abs/2406.15237}{{\ttfamily
  2406.15237}}].

\bibitem{Bevins2024}
H.T.J.~{Bevins}, S.~{Heimersheim}, I.~{Abril-Cabezas}, A.~{Fialkov}, E.~{de
  Lera Acedo}, W.~{Handley} et~al., \emph{{Joint analysis constraints on the
  physics of the first galaxies with low-frequency radio astronomy data}},
  \href{https://doi.org/10.1093/mnras/stad3194}{\emph{\mnras} {\bfseries 527}
  (2024) 813} [\href{https://arxiv.org/abs/2301.03298}{{\ttfamily
  2301.03298}}].

\bibitem{Pochinda2024}
S.~{Pochinda}, T.~{Gessey-Jones}, H.T.J.~{Bevins}, A.~{Fialkov},
  S.~{Heimersheim}, I.~{Abril-Cabezas} et~al., \emph{{Constraining the
  properties of Population III galaxies with multiwavelength observations}},
  \href{https://doi.org/10.1093/mnras/stae1185}{\emph{\mnras} {\bfseries 531}
  (2024) 1113} [\href{https://arxiv.org/abs/2312.08095}{{\ttfamily
  2312.08095}}].

\bibitem{Dhandha2025}
J.~{Dhandha}, A.~{Fialkov}, T.~{Gessey-Jones}, H.T.J.~{Bevins}, S.~{Tacchella},
  S.~{Pochinda} et~al., \emph{{Exploiting synergies between JWST and cosmic
  21-cm observations to uncover star formation in the early Universe}},
  \href{https://doi.org/10.1093/mnras/staf1359}{\emph{\mnras} {\bfseries 542}
  (2025) 2292} [\href{https://arxiv.org/abs/2503.21687}{{\ttfamily
  2503.21687}}].

\bibitem{Dhandha2025a}
J.~{Dhandha}, A.~{Fialkov}, T.~{Gessey-Jones}, H.T.J.~{Bevins}, S.~{Tacchella},
  S.~{Pochinda} et~al., \emph{{Narrowing the discovery space of the
  cosmological 21-cm signal using multi-wavelength constraints}},
  \href{https://doi.org/10.1093/mnras/staf1736}{\emph{\mnras} {\bfseries 544}
  (2025) 1608} [\href{https://arxiv.org/abs/2508.13761}{{\ttfamily
  2508.13761}}].

\bibitem{Davies2025}
J.E.~{Davies}, A.~{Mesinger} and S.G.~{Murray}, \emph{{Efficient simulation of
  discrete galaxy populations and associated radiation fields over the first
  billion years}},
  \href{https://doi.org/10.1051/0004-6361/202554951}{\emph{\aap} {\bfseries
  701} (2025) A236} [\href{https://arxiv.org/abs/2504.17254}{{\ttfamily
  2504.17254}}].

\bibitem{Semelin2007}
B.~{Semelin}, F.~{Combes} and S.~{Baek}, \emph{{Lyman-alpha radiative transfer
  during the epoch of reionization: contribution to 21-cm signal
  fluctuations}}, \href{https://doi.org/10.1051/0004-6361:20077965}{\emph{\aap}
  {\bfseries 474} (2007) 365}
  [\href{https://arxiv.org/abs/0707.2483}{{\ttfamily 0707.2483}}].

\bibitem{Baek2009}
S.~{Baek}, P.~{Di Matteo}, B.~{Semelin}, F.~{Combes} and Y.~{Revaz}, \emph{{The
  simulated 21 cm signal during the epoch of reionization: full modeling of the
  Ly-{\ensuremath{\alpha}} pumping}},
  \href{https://doi.org/10.1051/0004-6361:200810757}{\emph{\aap} {\bfseries
  495} (2009) 389} [\href{https://arxiv.org/abs/0808.0925}{{\ttfamily
  0808.0925}}].

\bibitem{Semelin2017}
B.~{Semelin}, E.~{Eames}, F.~{Bolgar} and M.~{Caillat}, \emph{{21SSD: a public
  data base of simulated 21-cm signals from the epoch of reionization}},
  \href{https://doi.org/10.1093/mnras/stx2274}{\emph{\mnras} {\bfseries 472}
  (2017) 4508} [\href{https://arxiv.org/abs/1707.02073}{{\ttfamily
  1707.02073}}].

\bibitem{21CMFAST}
A.~{Mesinger}, S.~{Furlanetto} and R.~{Cen}, \emph{{21CMFAST: a fast,
  seminumerical simulation of the high-redshift 21-cm signal}},
  \href{https://doi.org/10.1111/j.1365-2966.2010.17731.x}{\emph{\mnras}
  {\bfseries 411} (2011) 955}
  [\href{https://arxiv.org/abs/1003.3878}{{\ttfamily 1003.3878}}].

\bibitem{Santos2010}
M.G.~{Santos}, L.~{Ferramacho}, M.B.~{Silva}, A.~{Amblard} and A.~{Cooray},
  \emph{{Fast large volume simulations of the 21-cm signal from the
  reionization and pre-reionization epochs}},
  \href{https://doi.org/10.1111/j.1365-2966.2010.16898.x}{\emph{\mnras}
  {\bfseries 406} (2010) 2421}
  [\href{https://arxiv.org/abs/0911.2219}{{\ttfamily 0911.2219}}].

\bibitem{Fialkov2012}
A.~{Fialkov}, R.~{Barkana}, D.~{Tseliakhovich} and C.M.~{Hirata}, \emph{{Impact
  of the relative motion between the dark matter and baryons on the first
  stars: semi-analytical modelling}},
  \href{https://doi.org/10.1111/j.1365-2966.2012.21318.x}{\emph{\mnras}
  {\bfseries 424} (2012) 1335}
  [\href{https://arxiv.org/abs/1110.2111}{{\ttfamily 1110.2111}}].

\bibitem{Visbal2012}
E.~{Visbal}, R.~{Barkana}, A.~{Fialkov}, D.~{Tseliakhovich} and C.M.~{Hirata},
  \emph{{The signature of the first stars in atomic hydrogen at redshift 20}},
  \href{https://doi.org/10.1038/nature11177}{\emph{\nat} {\bfseries 487} (2012)
  70} [\href{https://arxiv.org/abs/1201.1005}{{\ttfamily 1201.1005}}].

\bibitem{Thomas2009}
R.M.~{Thomas}, S.~{Zaroubi}, B.~{Ciardi}, A.H.~{Pawlik}, P.~{Labropoulos},
  V.~{Jeli{\'c}} et~al., \emph{{Fast large-scale reionization simulations}},
  \href{https://doi.org/10.1111/j.1365-2966.2008.14206.x}{\emph{\mnras}
  {\bfseries 393} (2009) 32} [\href{https://arxiv.org/abs/0809.1326}{{\ttfamily
  0809.1326}}].

\bibitem{Ghara2018}
R.~{Ghara}, G.~{Mellema}, S.K.~{Giri}, T.R.~{Choudhury}, K.K.~{Datta} and
  S.~{Majumdar}, \emph{{Prediction of the 21-cm signal from reionization:
  comparison between 3D and 1D radiative transfer schemes}},
  \href{https://doi.org/10.1093/mnras/sty314}{\emph{\mnras} {\bfseries 476}
  (2018) 1741} [\href{https://arxiv.org/abs/1710.09397}{{\ttfamily
  1710.09397}}].

\bibitem{BEORN}
T.~{Schaeffer}, S.K.~{Giri} and A.~{Schneider}, \emph{{BEORN: a fast and
  flexible framework to simulate the epoch of reionization and cosmic dawn}},
  \href{https://doi.org/10.1093/mnras/stad2937}{\emph{\mnras} {\bfseries 526}
  (2023) 2942} [\href{https://arxiv.org/abs/2305.15466}{{\ttfamily
  2305.15466}}].

\bibitem{Mirocha14}
J.~{Mirocha}, \emph{{Decoding the X-ray properties of pre-reionization era
  sources}}, \href{https://doi.org/10.1093/mnras/stu1193}{\emph{\mnras}
  {\bfseries 443} (2014) 1211}
  [\href{https://arxiv.org/abs/1406.4120}{{\ttfamily 1406.4120}}].

\bibitem{ECHO21}
S.~{Mittal}, G.~{Kulkarni} and P.~{Sims}, \emph{{ECHO21: a tool for modelling
  global 21-cm signal from dark ages to reionization}},
  \href{https://doi.org/10.1093/rasti/rzag001}{\emph{RAS Techniques and
  Instruments} {\bfseries 5} (2026) rzag001}
  [\href{https://arxiv.org/abs/2503.11762}{{\ttfamily 2503.11762}}].

\bibitem{RS18}
J.~{Raste} and S.~{Sethi}, \emph{{An Analytic Formulation of the 21 cm Signal
  from the Early Phase of the Epoch of Reionization}},
  \href{https://doi.org/10.3847/1538-4357/aac2d8}{\emph{\apj} {\bfseries 860}
  (2018) 55}.

\bibitem{RS19}
J.~{Raste} and S.~{Sethi}, \emph{Analytic formulation of 21 cm signal from
  cosmic dawn: Ly$\alpha$ fluctuations},
  \href{https://doi.org/10.3847/1538-4357/ab13a6}{\emph{\apj} {\bfseries 876}
  (2019) 56}.

\bibitem{Zeus21}
J.B.~{Mu{\~n}oz}, \emph{{An effective model for the cosmic-dawn 21-cm signal}},
  \href{https://doi.org/10.1093/mnras/stad1512}{\emph{\mnras} {\bfseries 523}
  (2023) 2587} [\href{https://arxiv.org/abs/2302.08506}{{\ttfamily
  2302.08506}}].

\bibitem{SCRIPT}
T.R.~{Choudhury} and A.~{Paranjape}, \emph{{Photon number conservation and the
  large-scale 21 cm power spectrum in seminumerical models of reionization}},
  \href{https://doi.org/10.1093/mnras/sty2551}{\emph{\mnras} {\bfseries 481}
  (2018) 3821} [\href{https://arxiv.org/abs/1807.00836}{{\ttfamily
  1807.00836}}].

\bibitem{Choudhury:2025}
T.R.~{Choudhury} and A.~{Chakraborty}, \emph{{Capturing small-scale
  reionization physics: A sub-grid model for photon sinks with SCRIPT}},
  \href{https://doi.org/10.1088/1475-7516/2025/10/114}{\emph{\jcap} {\bfseries
  2025} (2025) 114} [\href{https://arxiv.org/abs/2504.03384}{{\ttfamily
  2504.03384}}].

\bibitem{Planck2018}
{Planck Collaboration}, \emph{{Planck 2018 results. VI. Cosmological
  parameters}}, \href{https://doi.org/10.1051/0004-6361/201833910}{\emph{\aap}
  {\bfseries 641} (2020) A6}
  [\href{https://arxiv.org/abs/1807.06209}{{\ttfamily 1807.06209}}].

\bibitem{Hahn2011}
O.~{Hahn} and T.~{Abel}, \emph{{Multi-scale initial conditions for cosmological
  simulations}},
  \href{https://doi.org/10.1111/j.1365-2966.2011.18820.x}{\emph{\mnras}
  {\bfseries 415} (2011) 2101}
  [\href{https://arxiv.org/abs/1103.6031}{{\ttfamily 1103.6031}}].

\bibitem{Sheth2002}
R.K.~{Sheth} and G.~{Tormen}, \emph{{An excursion set model of hierarchical
  clustering: ellipsoidal collapse and the moving barrier}},
  \href{https://doi.org/10.1046/j.1365-8711.2002.04950.x}{\emph{\mnras}
  {\bfseries 329} (2002) 61}
  [\href{https://arxiv.org/abs/astro-ph/0105113}{{\ttfamily
  astro-ph/0105113}}].

\bibitem{Pritchard:2006}
J.R.~Pritchard and S.R.~Furlanetto, \emph{Descending from on high: {L}yman
  series cascades and spin-kinetic temperature coupling in the 21\,cm line},
  \href{https://doi.org/10.1111/j.1365-2966.2006.10028.x}{\emph{Mon. Not. Roy.
  Astron. Soc.} {\bfseries 367} (2006) 1057}
  [\href{https://arxiv.org/abs/astro-ph/0508381}{{\ttfamily
  astro-ph/0508381}}].

\bibitem{CM04}
X.~{Chen} and J.~{Miralda-Escud{\'e}}, \emph{{The Spin-Kinetic Temperature
  Coupling and the Heating Rate due to Ly{$\alpha$} Scattering before
  Reionization: Predictions for 21 Centimeter Emission and Absorption}},
  \href{https://doi.org/10.1086/380829}{\emph{\apj} {\bfseries 602} (2004) 1}
  [\href{https://arxiv.org/abs/astro-ph/0303395}{{\ttfamily
  astro-ph/0303395}}].

\bibitem{RSS24}
J.~{Raste}, A.K.~{Sarkar} and S.K.~{Sethi}, \emph{{Thermal Evolution of the
  Intergalactic Medium due to Ly{\ensuremath{\alpha}} Photons during the Cosmic
  Dawn}}, \href{https://doi.org/10.3847/1538-4357/ad84ec}{\emph{\apj}
  {\bfseries 976} (2024) 236}
  [\href{https://arxiv.org/abs/2406.16542}{{\ttfamily 2406.16542}}].

\bibitem{RS26}
J.~{Raste} and S.K.~{Sethi}, \emph{{21 cm Signal from the Thermal Evolution of
  Ly{\ensuremath{\alpha}} during Cosmic Dawn}},
  \href{https://doi.org/10.3847/1538-4357/ae2029}{\emph{\apj} {\bfseries 996}
  (2026) 44} [\href{https://arxiv.org/abs/2506.12827}{{\ttfamily 2506.12827}}].

\bibitem{Wouthuysen1952}
S.A.~{Wouthuysen}, \emph{{On the excitation mechanism of the 21-cm
  (radio-frequency) interstellar hydrogen emission line.}},
  \href{https://doi.org/10.1086/106661}{\emph{\aj} {\bfseries 57} (1952) 31}.

\bibitem{Field1958}
G.B.~{Field}, \emph{{Excitation of the Hydrogen 21-CM Line}},
  \href{https://doi.org/10.1109/JRPROC.1958.286741}{\emph{Proceedings of the
  IRE} {\bfseries 46} (1958) 240}.

\bibitem{Mineo2012}
S.~{Mineo}, M.~{Gilfanov} and R.~{Sunyaev}, \emph{{X-ray emission from
  star-forming galaxies - I. High-mass X-ray binaries}},
  \href{https://doi.org/10.1111/j.1365-2966.2011.19862.x}{\emph{\mnras}
  {\bfseries 419} (2012) 2095}
  [\href{https://arxiv.org/abs/1105.4610}{{\ttfamily 1105.4610}}].

\bibitem{Pritchard2007}
J.R.~{Pritchard} and S.R.~{Furlanetto}, \emph{{21-cm fluctuations from
  inhomogeneous X-ray heating before reionization}},
  \href{https://doi.org/10.1111/j.1365-2966.2007.11519.x}{\emph{\mnras}
  {\bfseries 376} (2007) 1680}
  [\href{https://arxiv.org/abs/astro-ph/0607234}{{\ttfamily
  astro-ph/0607234}}].

\bibitem{1985ApJ...298..268S}
J.M.~{Shull} and M.E.~{van Steenberg}, \emph{{X-ray secondary heating and
  ionization in quasar emission-line clouds}},
  \href{https://doi.org/10.1086/163605}{\emph{\apj} {\bfseries 298} (1985)
  268}.

\bibitem{Heating2001}
A.~{Venkatesan}, M.L.~{Giroux} and J.M.~{Shull}, \emph{{Heating and Ionization
  of the Intergalactic Medium by an Early X-Ray Background}},
  \href{https://doi.org/10.1086/323691}{\emph{\apj} {\bfseries 563} (2001) 1}
  [\href{https://arxiv.org/abs/astro-ph/0108168}{{\ttfamily
  astro-ph/0108168}}].

\bibitem{Furlanetto:2010}
S.~Furlanetto and S.J.~Stoever, \emph{Secondary ionization and heating by fast
  electrons},
  \href{https://doi.org/10.1111/j.1365-2966.2010.16401.x}{\emph{Mon. Not. Roy.
  Astron. Soc.} {\bfseries 404} (2010) 1869}
  [\href{https://arxiv.org/abs/0910.4410}{{\ttfamily 0910.4410}}].

\bibitem{Bouwens2021}
R.J.~{Bouwens}, P.A.~{Oesch}, M.~{Stefanon}, G.~{Illingworth}, I.~{Labb{\'e}},
  N.~{Reddy} et~al., \emph{{New Determinations of the UV Luminosity Functions
  from z 9 to 2 Show a Remarkable Consistency with Halo Growth and a Constant
  Star Formation Efficiency}},
  \href{https://doi.org/10.3847/1538-3881/abf83e}{\emph{\aj} {\bfseries 162}
  (2021) 47} [\href{https://arxiv.org/abs/2102.07775}{{\ttfamily 2102.07775}}].

\bibitem{Bouwens2022}
R.J.~{Bouwens}, G.~{Illingworth}, R.S.~{Ellis}, P.~{Oesch} and M.~{Stefanon},
  \emph{{z 2-9 Galaxies Magnified by the Hubble Frontier Field Clusters. II.
  Luminosity Functions and Constraints on a Faint-end Turnover}},
  \href{https://doi.org/10.3847/1538-4357/ac86d1}{\emph{\apj} {\bfseries 940}
  (2022) 55} [\href{https://arxiv.org/abs/2205.11526}{{\ttfamily 2205.11526}}].

\bibitem{Donnan2024}
C.T.~{Donnan}, R.J.~{McLure}, J.S.~{Dunlop}, D.J.~{McLeod}, D.~{Magee},
  K.Z.~{Arellano-C{\'o}rdova} et~al., \emph{{JWST PRIMER: a new multifield
  determination of the evolving galaxy UV luminosity function at redshifts z =
  9 - 15}}, \href{https://doi.org/10.1093/mnras/stae2037}{\emph{\mnras}
  {\bfseries 533} (2024) 3222}
  [\href{https://arxiv.org/abs/2403.03171}{{\ttfamily 2403.03171}}].

\bibitem{Gaikwad2020}
P.~{Gaikwad}, M.~{Rauch}, M.G.~{Haehnelt}, E.~{Puchwein}, J.S.~{Bolton},
  L.C.~{Keating} et~al., \emph{{Probing the thermal state of the intergalactic
  medium at z > 5 with the transmission spikes in high-resolution Ly
  {\ensuremath{\alpha}} forest spectra}},
  \href{https://doi.org/10.1093/mnras/staa907}{\emph{\mnras} {\bfseries 494}
  (2020) 5091} [\href{https://arxiv.org/abs/2001.10018}{{\ttfamily
  2001.10018}}].

\bibitem{Zhu2023}
Y.~{Zhu}, G.D.~{Becker}, H.M.~{Christenson}, A.~{D'Aloisio}, S.E.I.~{Bosman},
  T.~{Bakx} et~al., \emph{{Probing Ultralate Reionization: Direct Measurements
  of the Mean Free Path over 5 < z < 6}},
  \href{https://doi.org/10.3847/1538-4357/aceef4}{\emph{\apj} {\bfseries 955}
  (2023) 115} [\href{https://arxiv.org/abs/2308.04614}{{\ttfamily
  2308.04614}}].

\bibitem{Wyithe2011}
J.S.B.~{Wyithe} and J.S.~{Bolton}, \emph{{Near-zone sizes and the rest-frame
  extreme ultraviolet spectral index of the highest redshift quasars}},
  \href{https://doi.org/10.1111/j.1365-2966.2010.18030.x}{\emph{\mnras}
  {\bfseries 412} (2011) 1926}
  [\href{https://arxiv.org/abs/1008.1107}{{\ttfamily 1008.1107}}].

\bibitem{Calverley2011}
A.P.~{Calverley}, G.D.~{Becker}, M.G.~{Haehnelt} and J.S.~{Bolton},
  \emph{{Measurements of the ultraviolet background at 4.6 < z < 6.4 using the
  quasar proximity effect}},
  \href{https://doi.org/10.1111/j.1365-2966.2010.18072.x}{\emph{\mnras}
  {\bfseries 412} (2011) 2543}
  [\href{https://arxiv.org/abs/1011.5850}{{\ttfamily 1011.5850}}].

\bibitem{DAloisio2018}
A.~{D'Aloisio}, M.~{McQuinn}, F.B.~{Davies} and S.R.~{Furlanetto}, \emph{{Large
  fluctuations in the high-redshift metagalactic ionizing background}},
  \href{https://doi.org/10.1093/mnras/stx2341}{\emph{\mnras} {\bfseries 473}
  (2018) 560} [\href{https://arxiv.org/abs/1611.02711}{{\ttfamily
  1611.02711}}].

\bibitem{Gaikwad2023}
P.~{Gaikwad}, M.G.~{Haehnelt}, F.B.~{Davies}, S.E.I.~{Bosman}, M.~{Molaro},
  G.~{Kulkarni} et~al., \emph{{Measuring the photoionization rate, neutral
  fraction, and mean free path of H I ionizing photons at 4.9
  {\ensuremath{\leq}} z {\ensuremath{\leq}} 6.0 from a large sample of XShooter
  and ESI spectra}},
  \href{https://doi.org/10.1093/mnras/stad2566}{\emph{\mnras} {\bfseries 525}
  (2023) 4093} [\href{https://arxiv.org/abs/2304.02038}{{\ttfamily
  2304.02038}}].

\bibitem{2026arXiv260630947C}
A.~{Chakraborty}, T.R.~{Choudhury}, K.K.~{Datta}, P.~{Dayal}, J.~{Dhandha},
  S.~{Gagnon-Hartman} et~al., \emph{{Square Kilometer Array Synergies for the
  Epoch of Reionization and Cosmic Dawn}},
  \href{https://doi.org/10.48550/arXiv.2606.30947}{\emph{arXiv e-prints} (2026)
  arXiv:2606.30947} [\href{https://arxiv.org/abs/2606.30947}{{\ttfamily
  2606.30947}}].

\end{thebibliography}
\end{document}